\documentclass[obeyspaces,
  journal=pasa,
  manuscript=research-paper, %% or "review"
  year=202X,
  volume=YY,]{cup-journal}
\usepackage{aas-macros}

\usepackage{amsmath}
\usepackage{amssymb,microtype,siunitx,booktabs}
\usepackage[skip=0.5ex]{subcaption}
\usepackage{url}
\usepackage{hyperref} %this package rather than url will create the actual active links
\usepackage{rotating}
\usepackage{tabularx}
\usepackage{orcidlink}
\usepackage{xfp}
\usepackage{braket}

\newcommand{\numdragnsG}{2687}
\newcommand{\dragndensityG}{\fpeval{round(\numdragnsG / 60,1)}} 

\newcommand{\numcompsG}{3055}
\newcommand{\compdensityG}{\fpeval{round(\numcompsG / 60,1)}}

\newcommand{\numrgcatG}{1823}
\newcommand{\rgcatdensityG}{\fpeval{round(\numrgcatG / 60,1)}}

\title{Extended Radio Galaxies in EMU: A Comparative Look at Source-Finding Techniques}

\author{Lachlan~J.~Barnes \orcidlink{0009-0008-8513-0619}}
\affiliation{School of Mathematical and Physical Sciences, Macquarie University, Balaclava Road, Sydney, NSW 2109, Australia}
\alsoaffiliation{Astrophysics and Space Technologies Research Centre, Macquarie University, Balaclava Road, Sydney, NSW 2109, Australia}
\email[Lachlan~J.~Barnes]{lachlan.barnes3@hdr.mq.edu.au}

\author{Andrew~M.~Hopkins \orcidlink{0000-0002-6097-2747}}
\affiliation{School of Mathematical and Physical Sciences, Macquarie University, Balaclava Road, Sydney, NSW 2109, Australia}
\alsoaffiliation{Astrophysics and Space Technologies Research Centre, Macquarie University, Balaclava Road, Sydney, NSW 2109, Australia}

\author{Yjan~Gordon}
\affiliation{Department of Physics, University of Wisconsin-Madison, 1150 University Avenue, Madison, WI 53706, USA}

\author{Nikhel~Gupta}
\affiliation{Australia Telescope National Facility, CSIRO, Space \& Astronomy, PO Box 1130, Bentley WA 6102, Australia}

\author{Gary~Segal \orcidlink{0000-0001-7165-8860}}
\affiliation{School of Mathematics and Physics, University of Queensland, St Lucia, Brisbane, QLD 4072, Australia}

\author{Heinz~Andernach}
\affiliation{Th\"uringer Landessternwarte, Sternwarte 5, D-07778 Tautenburg, Germany}
\alsoaffiliation{Permanent address: Universidad de Guanajuato, Departamento de Astronom\'{i}a, DCNE, Callej\'on de Jalisco s/n, Guanajuato 36023, Mexico}

\author{Michael~J.~I.~Brown \orcidlink{0000-0002-1207-9137}}
\affiliation{School of Physics \& Astronomy, Monash University, Clayton, VIC 3800, Australia}

\author{Duncan~Farrah \orcidlink{0000-0003-1748-2010}}
\affiliation{Department of Physics and Astronomy, University of Hawai`i at M\=anoa, 2505 Correa Rd., Honolulu, HI, 96822, USA}
\alsoaffiliation{Institute for Astronomy, University of Hawai`i, 2680 Woodlawn Dr., Honolulu, HI, 96822, USA}

\author{Stanislav~S.~Shabala}
\affiliation{School of Natural Sciences, University of Tasmania, Private Bag 37, Hobart, Tasmania 7001, Australia}

\author{Sarah~V.~White}
\affiliation{South African Astronomical Observatory (SAAO), PO Box 9, Observatory, 7935, South Africa}

\author{O.~Ivy~Wong}
\affiliation{Australia Telescope National Facility, CSIRO, Space \& Astronomy, PO Box 1130, Bentley WA 6102, Australia}
\alsoaffiliation{International Centre for Radio Astronomy Research (ICRAR), The University of Western Australia, Crawley, WA, Australia}

\received {dd Mmm YYYY}
\revised  {dd Mmm YYYY}
\accepted {dd Mmm YYYY}
\published{dd Mmm YYYY}

 %% First letter not capped

\begin{document}

\begin{abstract}
Extended radio sources present unique challenges for automated detection and classification in wide-field radio surveys. With current surveys such as the Evolutionary Map of the Universe (EMU), robust and scalable methods are essential to identify and catalogue these complex sources. We apply three automatic approaches to detect complex radio emission in EMU observations of the Galaxy And Mass Assembly (GAMA) 09 field (EMU-G09) in order to evaluate their relative strengths and limitations in preparation for large-scale application across future EMU data releases. These include DRAGN{\scriptsize HUNTER}, designed to detect likely DRAGNs (Double Radio sources associated with Active Galactic Nuclei) from a component catalogue; coarse-grained complexity, a metric designed to highlight regions of complex emission; and RG-CAT, a machine learning pipeline trained on radio sources identified in the EMU pilot survey. We find that together, the three methods recover nearly all extended sources in EMU-G09 but identify largely distinct, partially-overlapping subsets, with only $375$ sources identified by all finders. This demonstrates that a combination of complementary techniques will be required to achieve a complete census of extended radio sources in future large-scale surveys.
\end{abstract}

\section{Introduction} 

The study of extended radio sources has expanded rapidly over recent decades. Historically, radio telescopes did not have the sensitivity to detect faint sources, nor the resolution to detect extended sources, limiting the total number of extended radio sources in survey catalogues \citep[see][for a review]{norris_extragalactic_2017}. Large-scale surveys such as FIRST \citep{becker_first_1995, white_FIRST_1997}, NVSS \citep{condon_NVSS_1998} and LoTSS \citep[][]{shimwell_lotss1_2017, shimwell_lotss2_2019, shimwell_lotss3_2022} have increased the known number of radio sources dramatically (cataloguing $\sim\! 900,000$, $\sim\!1.8$ million and $\sim\!4.4$ million sources, respectively), many of which show extended emission ($\sim\!80\%$, $\sim\!16\%$, $\sim\!80\%$, respectively). Current radio telescope facilities, such as the Square Kilometre Array pathfinders \citep[see][for more details]{Norris_nextgen_2011, norris_pathfinders_2013}, and their respective large-scale surveys will further increase the number of known radio sources into the tens of millions. 

Double Radio sources associated with Active Galactic Nuclei \citep[DRAGNs;][]{leahy_dragns_1993} dominate the population of extended radio sources at high flux densities \citep[][]{norris_dragns_25}. These systems often show symmetric double-lobed structures with jets, hotspots, lobes, and compact cores coincident with a host galaxy \citep[e.g.,][]{laingbridle2014, hardcastle2020, ndungu_classification_2024}. 
Deeper and more sensitive observations reveal an increasingly broader diversity of AGN-powered structures that depart from traditional classifications, such as the traditional Fanaroff–Riley morphology types \citep[][]{fanaroff_morphology_1974}. These include X-shaped or winged radio galaxies \citep[e.g.,][]{Krishna_xshaped_2003}, remnant or dying radio sources with diffuse fading lobes \citep[e.g.,][]{brienza_renmant_2017}, and giant radio galaxies \citep[GRGs,][]{dabhade_discovery_2017, dabhade_grg_2020} extending over megaparsec scales. Extended radio emission is also not limited to AGN activity, and arises in a wide range of astrophysical environments. In nearby spiral galaxies, HII regions and supernova remnants contribute compact and filamentary radio features \citep[][]{condon_normalgals_1992}. On larger scales, galaxy groups and clusters host low surface brightness sources such as radio halos, and relics in the intracluster medium \citep[e.g.,][]{norris_evolutionary_2021, hopkins25}. Extended radio morphologies are diverse, often complex, and are shaped by various physical mechanisms beyond classical AGN jets.

As the diversity of extended radio sources becomes more apparent, the process of identifying, cataloguing, and classifying them has become increasingly challenging. Identifying the host galaxies of extended systems also adds further complexity, but is required to derive any physical parameters
of the sources (such as size or luminosity). Manual cross-matching is known to be labour-intensive and nonviable for large datasets \citep[e.g.][]{fan_xmatching_2015, sarah_g4JyI_2020, sarah_g4JyII_2020,  fan_xmatching_2020}. With modern surveys detecting millions of sources, automated approaches have become essential for scaling source identification and characterisation beyond what is feasible through human inspection alone.

Automatic source detection software such as \textit{Selavy} \citep[][]{Whiting_askapsource_2012}, \textit{PyBDSF} \citep[Python Blob Detector and Source Finder,][]{mohan_pybdsf_2015}, and \textit{CAESAR} \citep[][]{riggi_caeser_2016,riggi_caesar_2019} have been incorporated into radio telescope pipelines to accommodate the large influx of radio data. They have been shown to accurately identify and characterise compact sources \citep[e.g.,][]{hopkins_sourcefinding_2015,riggi_caesar_2021,Boyce_hydraI_2023, Boyce_hydraII_2023}, however there is mixed performance on extended sources. New tools have since been designed to identify and characterise extended radio emission, such as DRAGN{\footnotesize HUNTER} \citep[][]{Gordon_DRAGN_2023}, and coarse-grained complexity \citep[][]{Segal_Complex_2019}. Various machine learning approaches have also been developed \citep[e.g.,][]{ma_machine_2019, wu_radio_2019, galvin_cataloguing_2020, gupta_discovery_2022, riggi_maskcnn_2023, Gupta_rgcat_2024, alam2025_SOM}, which have been shown to be capable at handling the vast amounts of radio data being obtained and detecting extended sources. Complementary to these approaches, large-scale citizen-science projects such as Radio Galaxy Zoo \citep[RGZ,][]{banfield_rgz_2015, wong_rgz_2025} have shown that crowd-sourced human classifications remain highly effective for identifying complex extended systems, and can provide training sets to further advance machine learning approaches.

Identifying and characterising extended radio sources remains a central challenge in the era of large-scale radio continuum surveys. In this work, we perform a comparative analysis of three independent extended source-finding algorithms, applied to the same region. Rather than comparing to a predefined ground truth, we focus on understanding the relative behaviour of each source finder: the types of extended systems they successfully identify, the cases they miss, and the degree of overlap between their detections. This comparison provides practical insight into the strengths and limitations of current methods, and informs future development and application of extended source-finding techniques for current and upcoming wide-field surveys. This paper is structured as follows: \S\ref{data} describes the multi-wavelength data used in this work. \S\ref{sourcefinders} introduces and outlines the three extended source-finding algorithms. \S\ref{sourcefinder_comp} presents a comparative analysis of their performance and discusses implications for future surveys. Where relevant, we adopt a flat $\Lambda$CDM cosmology with $H_0 = 70$\,km\,s$^{-1}$, Mpc$^{-1}$, $\Omega_{\mathrm{M}} = 0.3$, $\Omega_\Lambda = 0.7$, and $\Omega_{\mathrm{k}} = 0$.

\section{Data} \label{data}

\subsection{GAMA}\label{gama_section}

Galaxy and Mass Assembly\footnote{\url{https://www.gama-survey.org/}} \citep[GAMA,][]{Driver_GAMA_2011, bellstedt_GAMA_2020, Driver_gamadr4_2022} is a multi-wavelength imaging and spectroscopic survey using the AAOmega multi-object spectrograph \citep[][]{smith_aaomega_2004} on the Anglo-Australian Telescope (AAT). The GAMA survey covers five separate fields (G02, G09, G12, G15, and G23), spanning a collective area of $286$ square degrees. Each region has a survey limit of $r < 19.8$\,mag, except for G23, which has a survey limit of $i<19.2$\,mag. Optical spectroscopy has been obtained for $\sim\!300,000$ galaxies, with a completeness of $\sim\!98\%$ at the main survey limit \citep{Driver_GAMA_2011, liske_gama_2015}. The spectra cover $3750$–$8850$\,\AA\ at a resolving power of $R \sim\!1300$, yielding typical redshift uncertainties of $50$–$100$\,km\,s$^{-1}$ \citep{hopkins_gama_2013, liske_gama_2015}. In addition to spectroscopy, GAMA incorporates multi-wavelength photometry, including ultraviolet (GALEX FUV, NUV), optical ($ugriz$ from SDSS/KiDS), near-infrared (UKIDSS/VISTA $YJHK_s$), mid-infrared (WISE), and far-infrared/sub-mm (Herschel PACS/SPIRE) imaging \citep{Driver_gama_2016, bellstedt_GAMA_2020}. The G09 field covers $\sim\!60$ square degrees (RA: $129^\circ$ to $141^\circ$, Dec: $-2^\circ$ to $+3^\circ$), with spectroscopic coverage matching the general GAMA limits and completeness described above.

We merge several public data products available through the GAMA Data Management Units\footnote{\url{https://www.gama-survey.org/dr4/schema/}} \citep[DMUs; see e.g.,][]{bellstedt_GAMA_2020, Driver_gamadr4_2022}. Specifically, we use the \texttt{gkvScienceCatv02}, which provides target selection, spectroscopic redshifts, and base photometry, \texttt{ProSpectAGNv02} for AGN-star formation decomposition from SED fitting, \texttt{StellarMassesGKVv24} for stellar mass and population parameters, and \texttt{gkvFarIRv03} for far-infrared fluxes. The merged dataset provides a unified set of spectroscopic, photometric, and derived astrophysical parameters for each galaxy. We then restricted this catalogue to the G09 region. The code used to construct this merged GAMA catalogue, hereafter referred to as our GAMA catalogue, is available from the authors upon request.

\subsection{EMU}

\begin{figure*}[htp!]
    \centering
    \includegraphics[width=\linewidth]{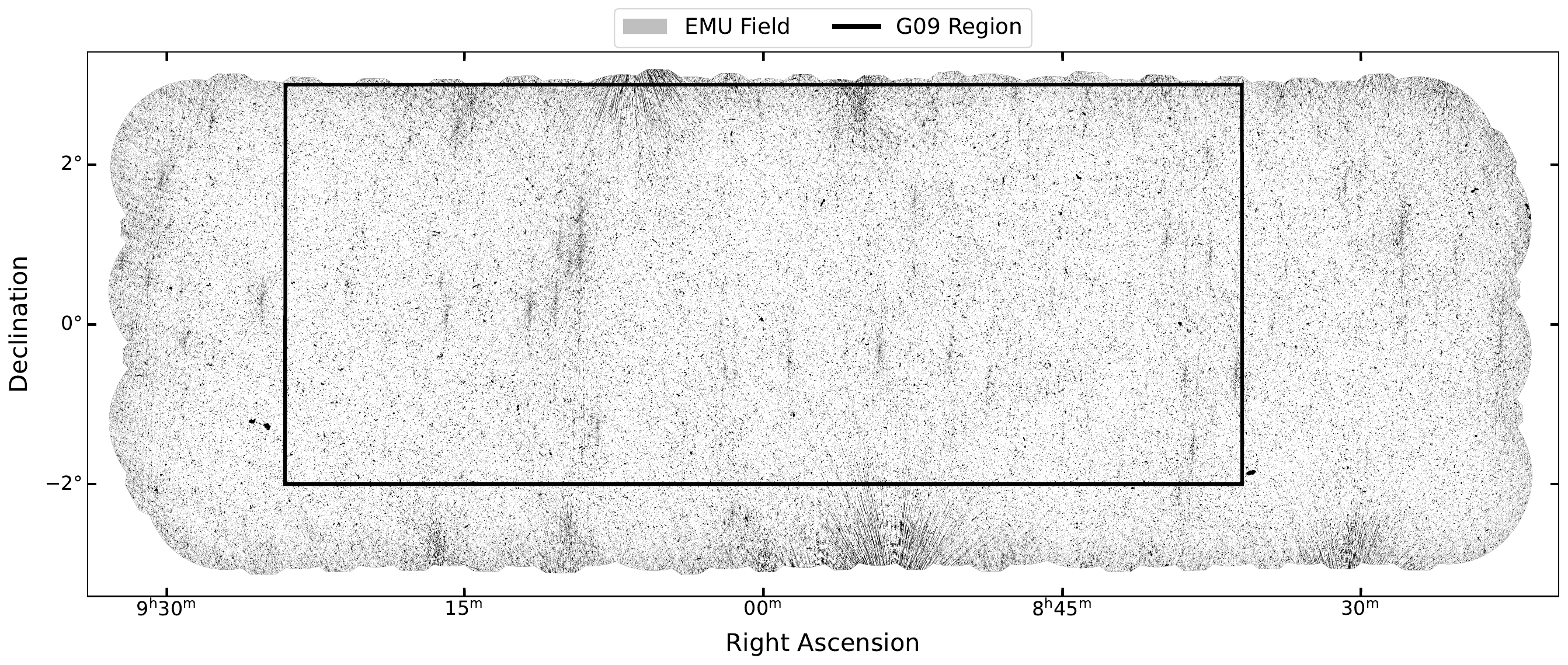}
    \caption{EMU mosaic constructed from the three EMU tiles that overlap the GAMA09 region with the black box indicating the GAMA09 footprint. The EMU observations are centred on a frequency of $944$ MHz, a native resolution of $11''\times13''$, and an rms sensitivity of $\sim\!25\,\mu$Jy beam$^{-1}$. These data constitute the EMU-G09 region used throughout this paper.}
    \label{emu_g09}
\end{figure*}

The Australian Square Kilometre Array Pathfinder (ASKAP) is a radio telescope facility located in Western Australia. It is comprised of $36$ dishes, each $12$\,m in diameter, located at the Murchison Radio-astronomy Observatory in Western Australia \citep{johnston_science_2007}. ASKAP has a phased-array feed (PAF) located at the focus of each antenna \citep{hotan_australian_2021} and an instantaneous field of view (FOV) of up to $30$ square degrees at $800$\,MHz \citep{norris_evolutionary_2021}, meaning that ASKAP is able to survey the sky approximately thirty times faster than traditional interferometers \citep{Norris_nextgen_2011}. The Evolutionary Map of the Universe\footnote{\url{https://emu-survey.org/}} \citep[EMU,][]{norris_emu_2011, norris_evolutionary_2021, hopkins25} is an ongoing wide-field radio continuum survey that is producing a highly sensitive atlas of the Southern Hemisphere using the ASKAP facility. The EMU survey is conducted at a frequency $\sim\!900$\,MHz, with a sensitivity of $20$-$30\,\mu$Jy\,beam$^{-1}$ at a resolution of $\sim\!15''$. It is anticipated that EMU will detect $\sim\!20$ million radio sources, with an estimate of hundreds of thousands showing well-resolved extended emission \citep[][]{hopkins25}. As there are no dedicated EMU observations on the G09 region, we combine three adjacent EMU tiles that overlap G09 (Figure~\ref{emu_g09}). These observations form part of the EMU main survey \citep[][]{hopkins25}. For all subsequent analysis, we restrict this combined field to the G09 footprint. The resulting EMU field, referred to hereafter as EMU-G09, has observational characteristics similar to those of the EMU Pilot Survey \citep[EMU-PS;][]{norris_evolutionary_2021}, with a central frequency of $944$ MHz, a native resolution of $11''\times13''$, and a sensitivity of $1$-$\sigma \approx 25\,\mu$Jy beam$^{-1}$.

\section{Source-Finding Approaches}\label{sourcefinders}
\subsection{Selavy}

\textit{Selavy} \citep[][]{Whiting_askapsource_2012}, derived from \textit{Duchamp} \citep[][]{whiting_duchamp_2012}, is the source-finding algorithm built into the ASKAP Science Data Processing software \citep[ASKAPsoft\footnote{\url{https://www.atnf.csiro.au/data-software/software/}},][]{whiting_pipeline_2017, norris_evolutionary_2021}. \textit{Selavy} identifies groups of pixels with flux densities above a threshold \citep[typically $>3\sigma$ of the local rms of a radio image, see \S3.2 of][ for more detail]{Whiting_askapsource_2012}, which are referred to as `islands'. The \textit{Selavy} island catalogue contains the sizes, positions and flux densities of each island. Gaussian components are then fit to the emission peaks of each island, producing a separate component catalogue \citep[see][]{norris_evolutionary_2021}. \textit{Selavy} identified a total of $34685$ islands in EMU-G09, which were fit with $37838$ components, with approximately $91.1\%$ of islands fit by a single component. While \textit{Selavy} is well-suited to detecting compact radio sources, it is not designed to reliably catalogue extended sources, which can appear as multiple distinct regions on the sky (e.g., two lobes and a core). The island and component catalogues can, however, provide a useful foundation for higher-level algorithms designed to detect more complex emission. Here, we apply three such extended source-finders to explore the range of complex extended radio sources present in EMU-G09. The source finders we focus on in this work are DRAGN{\footnotesize HUNTER} \citep[][]{Gordon_DRAGN_2023}, coarse-grained complexity \citep[][]{Segal_Complex_2019,segal_complexity_2023}, and RG-CAT \citep[][]{Gupta_rgcat_2024}. We describe these source finders, and their implementation, in more detail in subsequent sections.

\subsection{DRAGN{\footnotesize HUNTER}}

DRAGN{\footnotesize HUNTER}\footnote{\url{https://github.com/ygordon/dragn_hunter/tree/master/dragnhunter}} \citep[\textit{DH} hereafter,][]{Gordon_DRAGN_2023} is a source-finding script designed to identify likely DRAGNs, based on an input catalogue of detected radio components. This script was originally designed to be applied to data from the Karl G. Jansky Very Large Array (VLA) Sky Survey \citep[VLASS;][]{lacy_karl_2020}. The \textit{DH} script identifies candidate radio galaxies by searching for nearest-neighbour pairs of extended radio components from an input component catalogue. We make use of the \textit{Selavy} component catalogue of EMU-G09 for this purpose.

The \textit{DH} script requires input parameters defining the minimum component flux density and a minimum angular size. We adopt the 10th percentile (P10) of peak flux density as a conservative threshold to exclude the faintest components, which are more likely to be affected by noise and have poorly constrained fluxes. The P10 value corresponds to a minimum peak flux density of $0.20\,$mJy, roughly $8\sigma$ of the EMU sensitivity ($25\,\mu$Jy\,beam$^{-1}$). 

We find that $\sim\!7\%$ of the \textit{Selavy} components are unresolved with a deconvolved major axis, $\Psi$, indistinguishable from zero and an additional $\sim\!9\%$ with $\Psi<2''$. To exclude the smallest and potentially unreliable components while retaining the majority of sources, we again adopt the 10th percentile (P10) threshold of the range of $\Psi$ values in our sample. The P10 size threshold of corresponds to $3.56''$, and was determined after excluding unresolved components, ensuring that only genuinely resolved components contribute to the selection candidate pairs. Using these values, we identify a total $19320$ component pairs from the EMU-G09 \textit{Selavy} component catalogue. A given component may appear in more than one pair if it is the nearest neighbour of multiple candidates. In such cases, the pair with the smallest angular separation is flagged as the preferred pair. This results in a sample of $9515$ preferred pairs.

These preferred component pairs are further refined into likely DRAGNs by considering both the angular separation between components and the mean misalignment between components. \citet{Gordon_DRAGN_2023} define the mean misalignment as:

\begin{equation}
    \Delta \theta_{\text{mean}} = \frac{\Delta \theta_1 + \Delta \theta_2}{2}~,
\end{equation}\label{mean_misalign}

\noindent where $\Delta \theta_n$ is the relative misalignment (between $0^\circ$ and $90^\circ$) of the position angle of component $n$, $\theta_n$, relative to the position angle of the pair, $\theta_{\text{pair}}$ \citep[see Figure~4 in][]{Gordon_DRAGN_2023}, given by:

\begin{equation}
    \Delta \theta_n = |\theta_n - \theta_{\text{pair}} |~.
\end{equation}

\begin{figure}[h!]
    \centering
    \includegraphics[width=\linewidth]{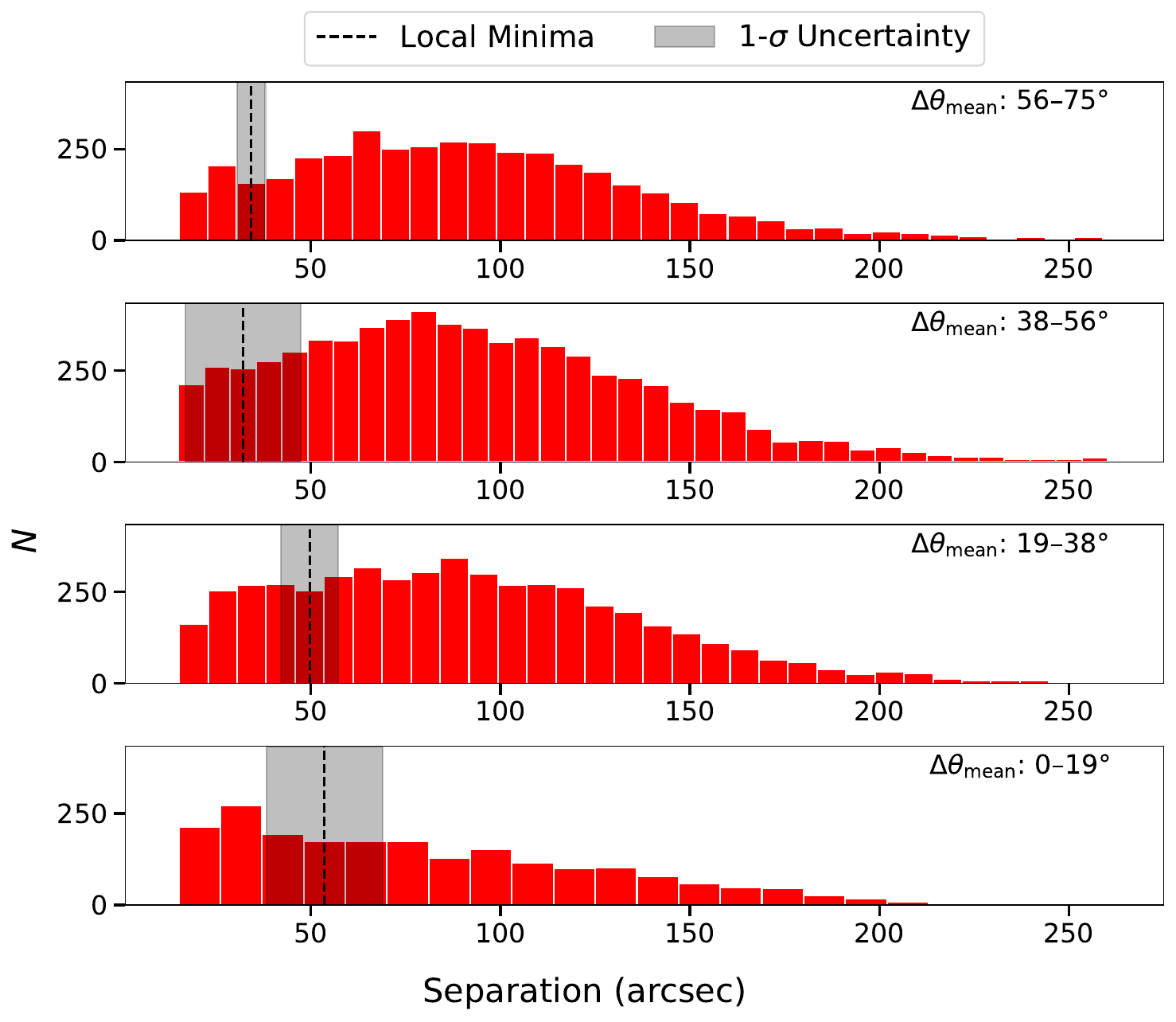}
    \caption{Distributions of angular separation of identified pairs in EMU-G09 (red bars) in bins of mean misalignment. Black lines mark the local minima, with the shaded region showing the $1\sigma$ uncertainty of their positions.}
    \label{antimode_hist}
\end{figure}

\noindent We bin the identified pairs in mean misalignment and measure the local minima of angular separation in each bin (Figure~\ref{antimode_hist}), to define where genuine multi-component sources decline and chance pairings begin to dominate. At small angular separations, pairs of components can generally be accepted as belonging to the same source since the probability of a chance association at small separations is low, even if the position angles of the pairs are misaligned. Conversely, for widely separated components, reliable associations preferentially occur when the components are well aligned, as larger separations increase the likelihood of random pairings. Conversely, for widely separated components, reliable associations are found when the components are well aligned, as larger separations increase the likelihood of random pairings. 

To refine the identified pairs into likely DRAGNs, we derive an upper limit in mean misalignment as a function of angular separation:

\begin{equation}
\Delta \theta_{\text{mean, EMU}} < -213.45 \, \text{log}_{10}(d) + 389.84~,
\label{EMU_antimode_fit_eqn}
\end{equation}

\noindent where $d$ is the angular separation in arcseconds. Only pairs with $d > 15''$ are considered, corresponding roughly to separations of at least one EMU beam. Pairs below this threshold may belong to a single \textit{Selavy} island or a marginally resolved source (see \ref{low_pair_sep}). 

\begin{figure}[h!]
    \centering
    \includegraphics[width=\linewidth]{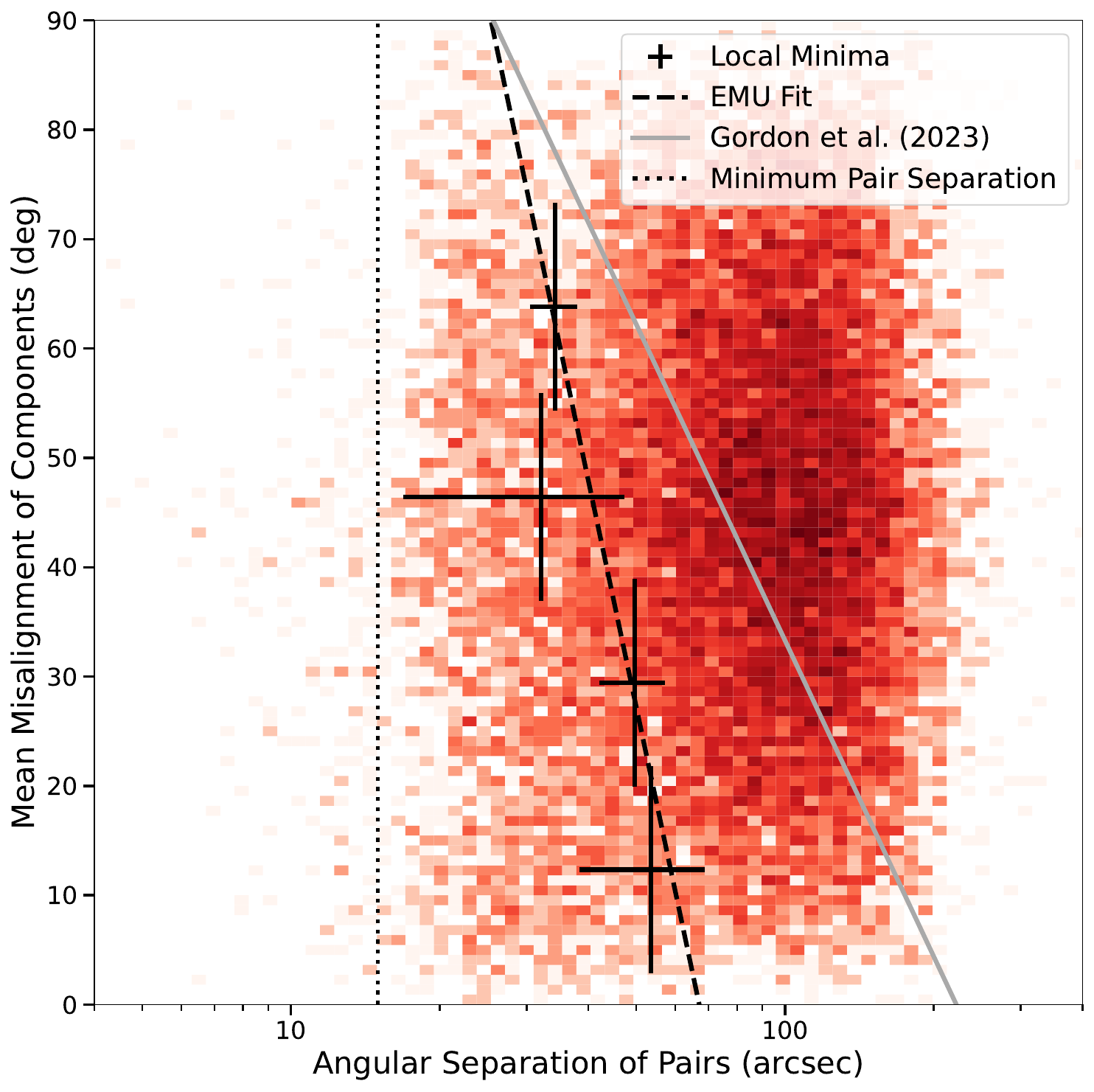}
    \caption{Mean misalignment and angular separation for candidate lobe pairs in EMU-G09. Black crosses show the local minima of pair separation in bins of mean misalignment, with vertical bars indicating bin size and horizontal bars the $1\sigma$ uncertainty. The black dashed and grey solid lines show the derived upper limits for likely DRAGNs in EMU and VLASS, respectively, reflecting their different observational parameters. The black dotted line represents the minimum pair separation ($15''$) used to select DRAGNs in this work.}
    \label{DRAGNHunter_pairs_antimodes}
\end{figure}

Figure~\ref{DRAGNHunter_pairs_antimodes} shows the identified pairs and the derived EMU upper limit, with the VLASS limit for comparison \citep{Gordon_DRAGN_2023}. The cluster of pairs at $\sim\!100''$ is likely dominated by chance associations, while pairs at smaller separations are more likely to be genuine multi-component sources. Comparing the EMU limit to the VLASS relation, $\Delta \theta_{\text{mean, VLASS}} < -96.01\ \text{log}_{10}(d) + 225.32$, we find that the EMU relation has a steeper slope and higher intercept, likely due to differences in resolution and observing frequency ($\sim\!15''$ vs $\sim\!2.5''$, and $944$\,MHz vs $\sim\!3$\,GHz, respectively). Using Equation~\ref{EMU_antimode_fit_eqn}, we identify $\numdragnsG$ likely DRAGNs from the $9515$ preferred candidate pairs in EMU-G09.

The \textit{DH} script estimates the total flux density, $S$, defined by \citet{Gordon_DRAGN_2023} as the sum of the constituent components of the DRAGN (i.e., both lobes and the core). Note that compact components ($S>0.20,$mJy and $\Psi<2''$), which may represent radio cores, are flagged and excluded from the initial lobe–lobe pair-finding stage, such that only extended components are considered when forming candidate pairs. To attempt to recover the missed cores, a search within $30''$ or half the pair separation (whichever is smaller) of the central position of the likely DRAGNs is conducted. For most sources however, no core is identified, and the total flux density corresponds to just the sum of both lobes. \textit{DH} also estimates the largest angular size (LAS) of the DRAGN. Here LAS is defined as the distance between the catalogue positions of the pair components, plus the semi-major axis lengths of each lobe. 

\begin{figure}[h!]
    \centering
    \includegraphics[width=\linewidth]{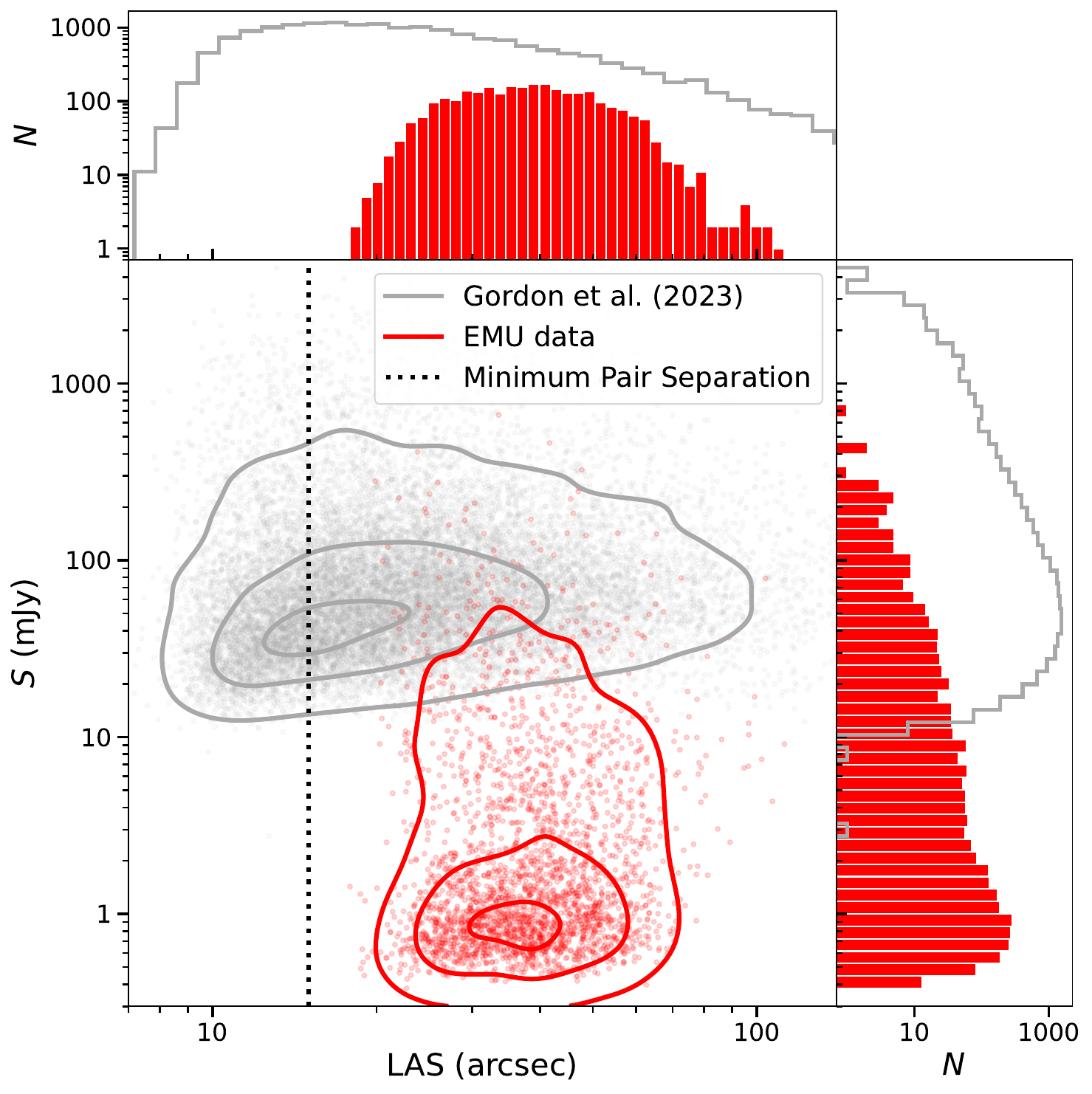}
    \caption{Distributions of LAS and integrated flux density ($S$) for likely DRAGNs in the EMU G09 field. Density contours of the distributions of LAS and $S$ for EMU DRAGNs (red) for the VLASS DRAGNs (grey). Contours contain $90\%$, $50\%$, and $10\%$ of the data points. The black dotted line represents the minimum pair separation ($15''$) used to select DRAGNs in this work. The difference between these distributions likely arises from the observational differences between EMU and VLASS.}
    \label{DRAGNHunter_LAS_Flux}
\end{figure}

The distributions of $S$ and LAS are shown in Figure~\ref{DRAGNHunter_LAS_Flux}. \citet{Gordon_DRAGN_2023} found that larger DRAGNs were generally brighter, however we do not explicitly see that trend in our EMU sample. The sources identified by \textit{DH} in EMU-G09 show little to no correlation between size and brightness. The VLASS DRAGNs also occupy a different region of this parameter space, though this is partly a consequence of the differing minimum component–separation thresholds applied to each survey. In EMU we exclude systems with separations below $15''$, whereas VLASS can identify structures down to $6''$. This means the EMU DRAGNs appear larger in Figure~\ref{DRAGNHunter_LAS_Flux} even though this is mostly a selection effect. As shown in the LAS histogram (top panel), the VLASS sample extends to larger angular sizes than our EMU sample. In addition to resolution effects, differences in survey sensitivity also contribute to the separation of the two samples in this parameter space. The significantly better sensitivity of EMU relative to VLASS means that the EMU-G09 sources are at systematically lower flux densities than VLASS sources.

\subsection{Coarse-Grained Complexity}\label{cg_comp_section}

\begin{figure}[h!]
    \centering
    \includegraphics[width=0.8\linewidth]{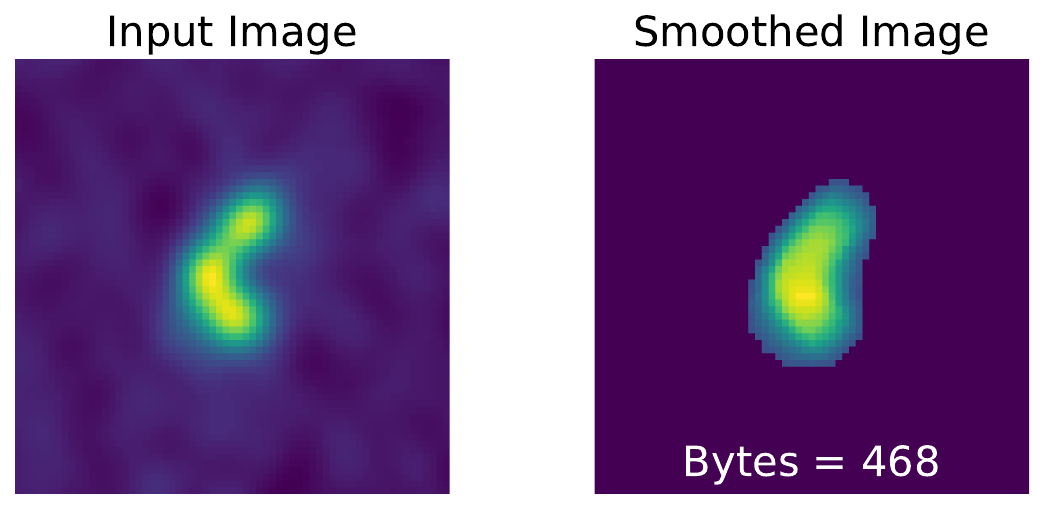}
    \caption{Illustrative example of the coarse-grained complexity process. An input image (left) is smoothed by a Gaussian filter to produce the smoothed image (right) which is then compressed using \texttt{gzip}. The byte length is then used as a proxy for apparent complexity of the radio source. The images here are $64\times64$ pixels, corresponding to an angular size of $128''\times128''$.}
    \label{cg_complexity_process}
\end{figure}

The coarse-grained complexity measure \citep[CG-Complexity,][]{Segal_Complex_2019, segal_complexity_2023} is a tool that can be applied to large datasets to identify morphologically complex sources in an efficient and automatic way. Coarse-grained complexity is an approximation of the Kolmogorov complexity \citep[see][for more details]{aaronson_complex_2014}. The CG-Complexity tool operates by first applying a smoothing kernel to an image (e.g., Gaussian or median filter), then compresses the smoothed image using \texttt{gzip} \citep[][]{levine_gzip_2012}, treating the byte length as an upper limit of the Kolmogorov complexity (Figure~\ref{cg_complexity_process}).

\citet{Segal_Complex_2019} and \citet{segal_complexity_2023} use cutouts of $64\times64$ and $256\times256$ pixels, respectively, and employ a blind search to identify complex regions of emission. We do not use a blind approach in this work. We instead create cutouts centred on the right ascension (RA) and declination (Dec) of each \textit{Selavy} island in EMU-G09. This distinction means we identify complex islands rather than complex regions, and therefore retain a dependence on the \textit{Selavy} source detections. When testing the larger cutout size, we found that multiple nearby islands were often included in a single cutout, resulting in repeated or correlated complexity estimates. We therefore adopt the smaller $64\times64$ cutouts, to minimise overlap between adjacent islands. While this means we may not capture the full extent of sources larger than $\sim\!128''$ (since one pixel is $2''\times2''$), we expect the obtained complexity values to more accurately represent the apparent complexity of each island. The smaller cutout size also significantly decreases the computation time when applying the coarse-grained complexity calculation to cutouts of the \textit{Selavy} islands. 

A manual inspection indicated that many cutouts were contaminated by bright nearby sources  towards the edges of the cutouts. Such contamination can artificially increase the measured apparent complexity. To identify and remove these cases systematically, we applied a central crop of each cutout ($32\times32$ pixels) and compared the peak pixel intensity within the crop to that of the full $64\times64$ cutout. Instances where the peak intensity in the full cutout was higher than the central crop were flagged as contaminated. This ensures that, in regions with multiple islands (where multiple cutouts may have very similar complexity), cutouts centred on fainter islands are discarded while the brightest island is retained. We identified $10220$ ($29.47\%$) contaminated cutouts, which were excluded from further analysis, leaving $24465$ cutouts of the \textit{Selavy} islands in our sample.

\begin{figure}[h!]
    \centering
    \includegraphics[width=\linewidth]{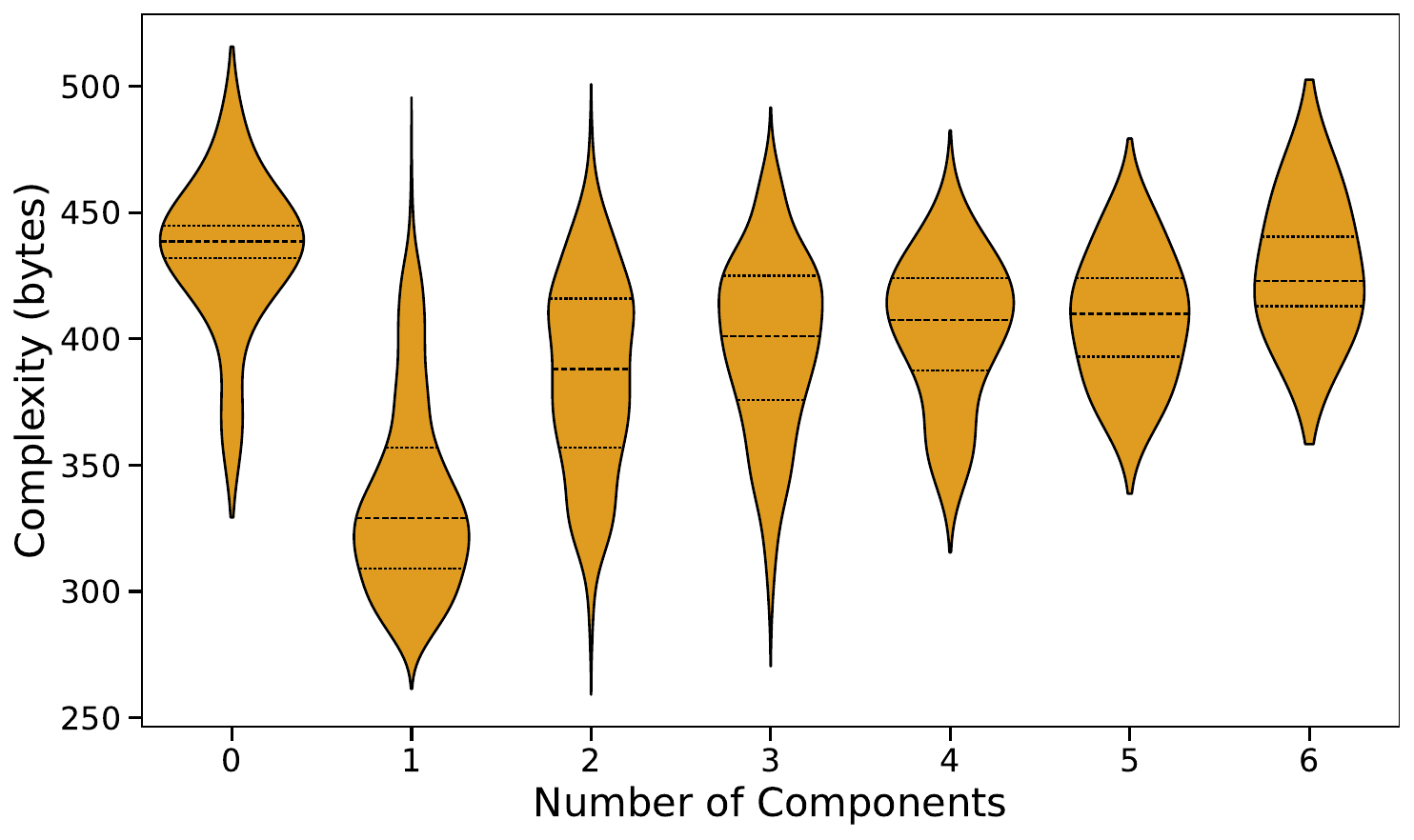}
    \caption{Source complexity as a function of the number of components fit to each \textit{Selavy} island, with quartiles marked inside. Note the general trend of increasing complexity with the number of components.}
    \label{comps_vs_complexity}
\end{figure}

The \textit{Selavy} island catalogue also reports the number of Gaussian components fit to each island. We examine how this relates to the CG-Complexity measure in Figure~\ref{comps_vs_complexity}. To quantify the relationship, we compute a Spearman rank correlation \citep[][]{siegel_nonparametric_1988, conover1999practical} using only sources with two or more components, since islands with zero or one component clearly deviate from any linear trend. We obtain a correlation coefficient of $0.281$ with a significance of $p = 8.2\times10^{-44}$. Although statistically significant, this weak correlation indicates that the number of fitted components is a much poorer proxy for apparent source complexity than the CG-Complexity metric. We discuss the zero-component islands in more detail in Section~\ref{0_comp_islands}.

\begin{figure}[h!]
    \centering
    \includegraphics[width=\linewidth]{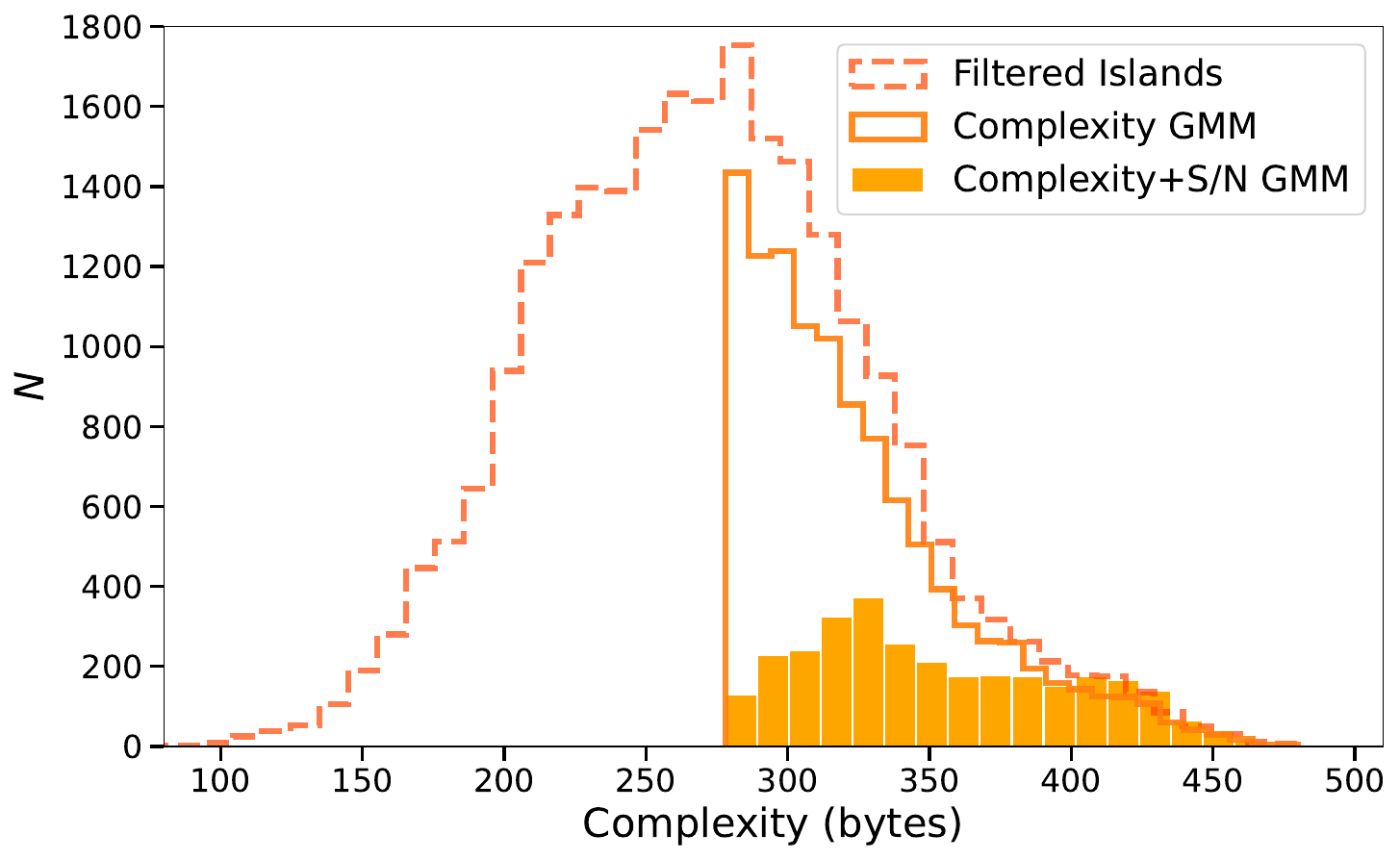}
    \caption{Comparison of distributions of the coarse-grained complexity values for all islands (dark orange dashed line), the complexity-only GMM (orange line), and the complexity$+$S/N GMM (light orange solid bars).}
    \label{cg_complexity_distributions}
\end{figure}

Following \citet{Segal_Complex_2019}, we use Gaussian mixture modelling (GMM) to create a subset of sources with significantly high complexity. We fit a two-component GMM to the complexity values for each of our islands from the \textit{Selavy} catalogue. On the cluster with the highest mean complexity, we then fit another two-component GMM using both the complexity values and signal-to-noise (S/N). Here we adopt a different definition for S/N than \citet{Segal_Complex_2019}, defining it as the peak flux over the background noise (both calculated by \textit{Selavy}) of each island. The distributions of complexities for each subset are highlighted in Figure~\ref{cg_complexity_distributions}. After fitting both GMMs, we produce a final dataset of $\numcompsG$ significantly complex sources with a minimum complexity of $278$ bytes. This value is comparable to the $\sim\!300$ byte threshold determined by \citet{Segal_Complex_2019} for significantly complex sources.

\begin{figure}[h!]
    \centering
    \begin{subfigure}{\linewidth}
       \centering
       \includegraphics[width=0.95\linewidth]{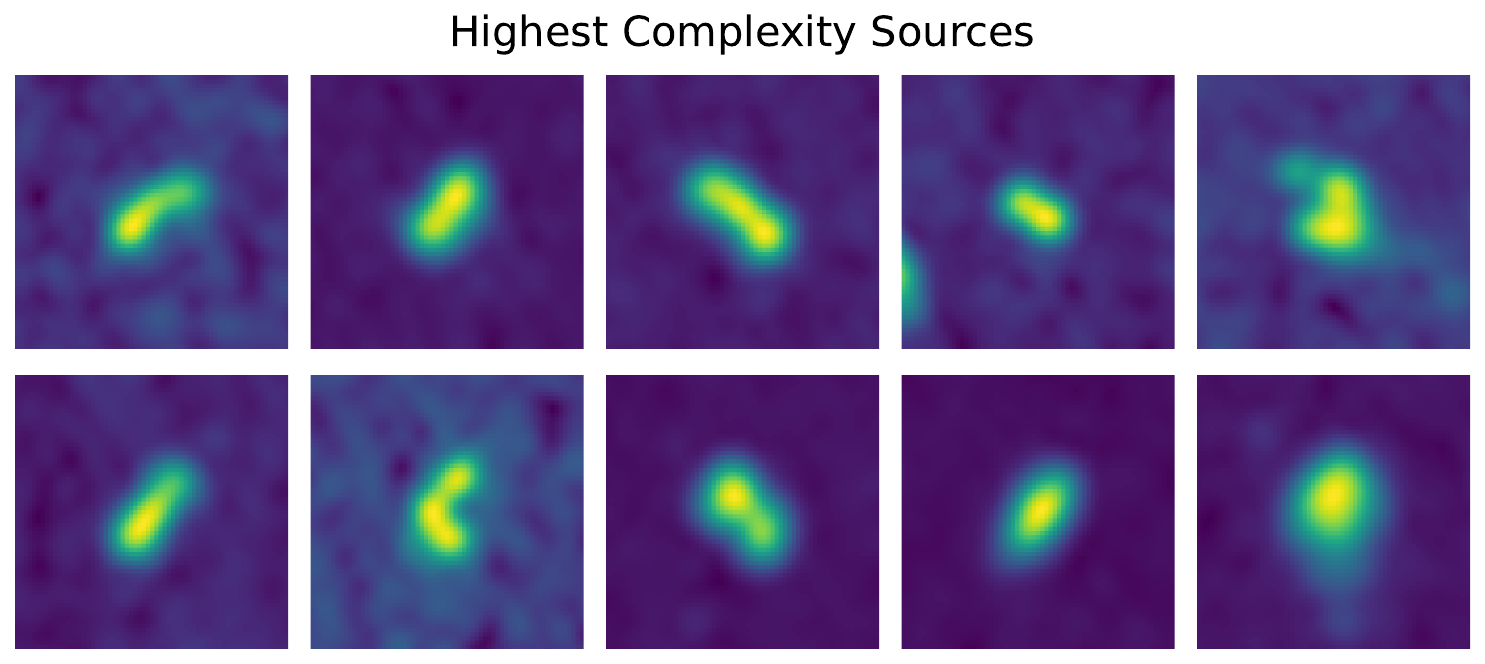}
    \end{subfigure}
    \begin{subfigure}{\linewidth}
        \centering
        \includegraphics[width=0.95\linewidth]{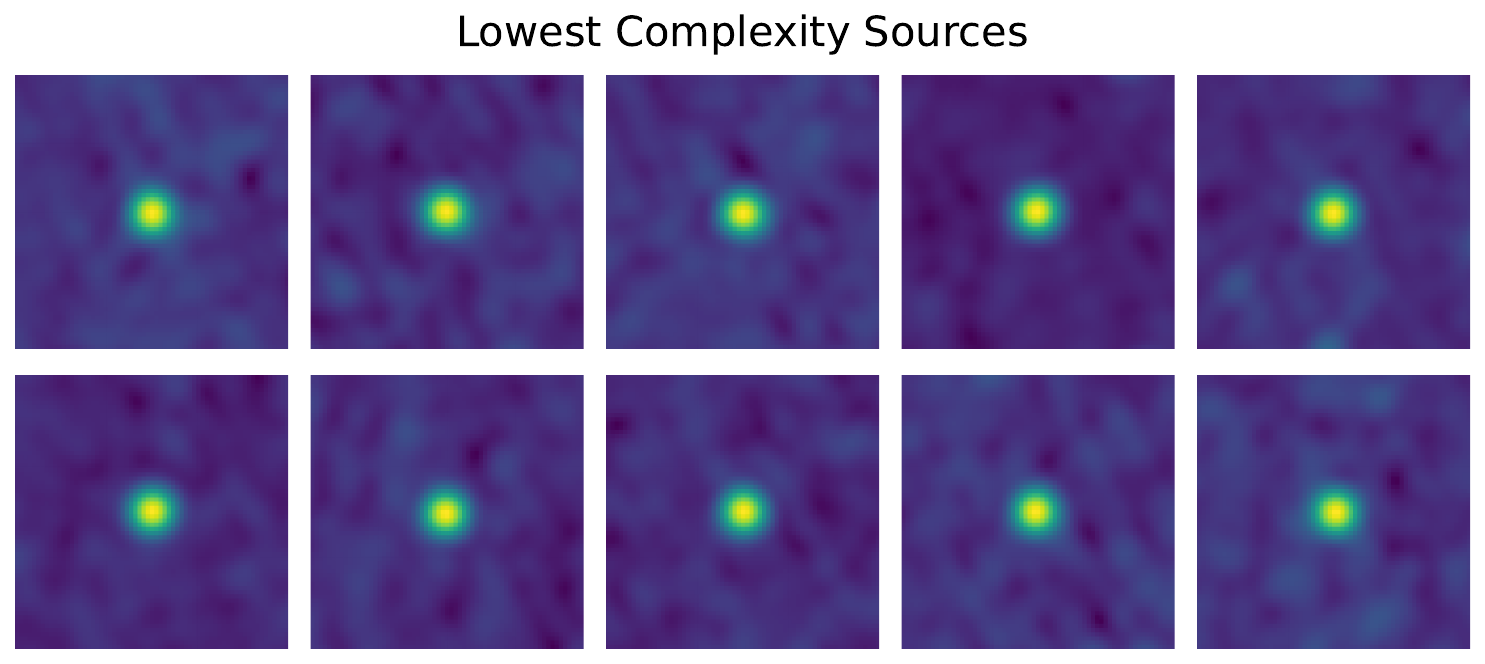}
    \end{subfigure}
    \caption{Examples of the ten most (top) and least (bottom) complex sources in the significantly complex dataset. The most complex sources display extended, multi-component, or diffuse morphologies, while the least complex examples are dominated by compact, unresolved sources where apparent complexity likely arises from background noise or faint nearby emission. All cutouts are $64\times64$ pixels, corresponding to an angular size of $128''\times128''$.}
    \label{high_low_complexity}
\end{figure}

Figure~\ref{high_low_complexity} presents examples of the ten most and least complex sources in the significantly complex sample. The highest-complexity sources encompass a wide variety of morphologies, including radio doubles, triples, slightly bent or asymmetric sources, and diffuse, irregular emission. Extended structure or multiple bright regions clearly contribute to high apparent complexity. In contrast, the lowest-complexity examples are visually compact unresolved point-like sources. Their presence in the significantly complex sample is somewhat unexpected. Even after smoothing, small-scale noise, sharp intensity gradients in bright unresolved sources, or minor residual artefacts may inflate the compressed byte-length of a cutout. Their inclusion likely reflects algorithmic sensitivity to high S/N gradients or subtle pixel-level variations, rather than genuine morphological complexity. We explore these `complex' point sources further in \ref{CG_complex_points}.

\subsection{RG-CAT}\label{rgcat_sec}

RG-CAT \citep[][]{Gupta_rgcat_2024} is a machine-learning (ML) detection pipeline designed to construct catalogues of radio sources (both extended and compact) from wide-field radio surveys. It leverages computer-vision methods \citep[e.g.,][]{gupta_deep_2023} to identify radio emission, associate related components, and locate potential infrared host galaxies within a unified framework. The model was trained on an extension of the RadioGalaxyNET dataset \citep[][]{gupta_radiogalnet_2024}, containing over $5000$ annotated radio galaxies and their infrared counterparts from the EMU Pilot Survey \citep[][]{norris_dragns_25}. Each $8'\times8'$ cutout in the training set includes detailed annotations: radio morphological classifications (e.g. FR-I, FR-II, compact, or rare types), bounding box parameters, segmentation masks for the radio emission, and host-galaxy positions obtained via infrared cross-matching.

\begin{figure}[h!]
    \centering
    \includegraphics[width=\linewidth]{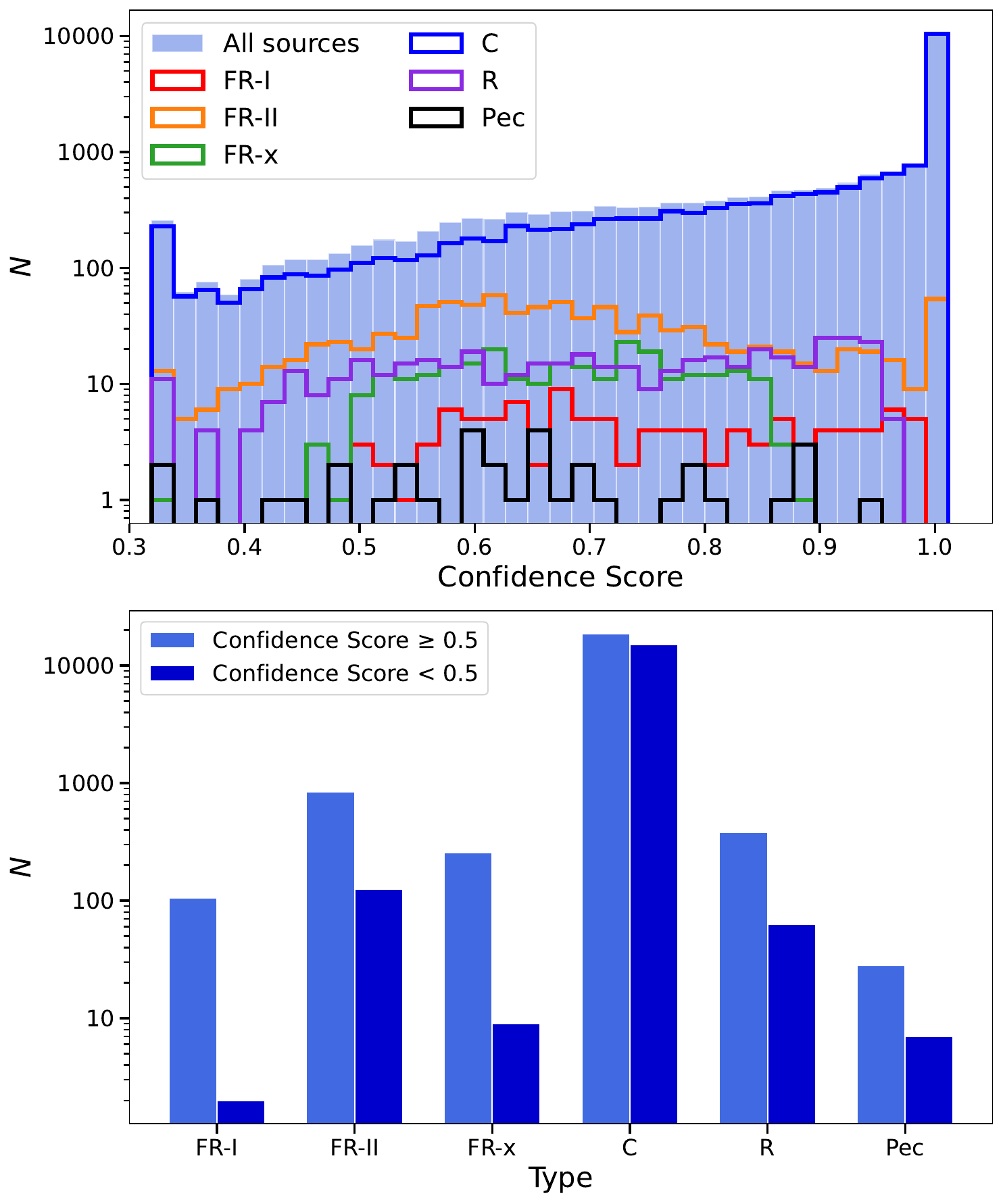}
    \caption{\textbf{Top panel:} Distributions of all sources (solid bars) and subsets of different classification types (coloured steps) based on the confidence score. \textbf{Bottom panel:} Number of different morphology types with confidence scores predicted by Gal-DINO above and below 0.5 (light blue and dark blue bars, respectively).}
    \label{rgcat_morphs_count}
\end{figure}

Gal-DINO \citep[as described in][]{gupta_radiogalnet_2024} was trained on the RG-CAT dataset to simultaneously predict bounding boxes for radio galaxies and potential keypoint positions of their infrared hosts, where a keypoint in ML refers to a specific point or landmark in images. In this work, we apply the Gal-DINO model to the EMU G09 field, without retraining, to produce a catalogue of candidate extended radio galaxies. $35533$ sources were identified in EMU-G09, with $\numrgcatG$ being extended. The lower panel of Figure~\ref{rgcat_morphs_count} illustrates the distribution of galaxies across the six prediction class types. RG-CAT identifies $33710$ compact sources (C), $108$ FR-I, $969$ FR-II, $265$ FR-x  (uncertain whether FR-I or FR-II), $446$ resolved (R), and $35$ peculiar morphologies (Pec). We will focus specifically on the $\numrgcatG$ extended sources, the $33710$ compact sources identified by RG-CAT are not used in this work. 

We show a comparison with the sources (and source types) of our data and the catalogue produced by \citet{Gupta_rgcat_2024} for the EMU-PS in Table~\ref{rgcat_g09_comparison}. The relative fractions of each morphological class in the G09 field are broadly consistent with those reported by \citet{Gupta_rgcat_2024}, suggesting that the RG-CAT model performs similarly across different sky regions. Both datasets are dominated by FR-II sources, which comprise roughly half of all extended detections, followed by smaller fractions of resolved (R) and FR-x morphologies. The proportions of FR-I and peculiar sources are also comparable between samples. Minor deviations, such as the slightly higher fraction of resolved sources and the lower fraction of peculiar morphologies in G09, likely reflect field-to-field variations in source density and morphology between the EMU-PS and G09 regions.

\begin{table}[h!]
\centering
\caption{Comparison of RG-CAT morphological class distributions between EMU-G09 sources and EMU-PS sources \citep{Gupta_rgcat_2024}, with fractions given relative to the total number of extended sources identified in each dataset.}
\begin{tabular}{lcccc}
\toprule
 & \multicolumn{2}{c}{EMU-G09} & \multicolumn{2}{c}{EMU-PS} \\
\cmidrule(lr){2-3} \cmidrule(lr){4-5}
 Type & Count & Fraction (\%) & Count & Fraction (\%) \\
\midrule
FR-I          & $108$  & $5.9$  & $582$   & $5.6$  \\
FR-II         & $969$ & $53.2$ & $5602$  & $53.8$ \\
FR-x         & $265$  & $14.5$ & $1494$  & $14.3$ \\
R            & $446$  & $24.5$ & $2375$  & $22.8$ \\
Pec          & $35$   & $1.9$  & $361$   & $3.5$  \\
\midrule
Total & $\numrgcatG$ & $100$ & $10414$ & $100$ \\
\bottomrule
\end{tabular}
\label{rgcat_g09_comparison}
\end{table}

\section{Source-Finder Comparison}\label{sourcefinder_comp}

\subsection{Relative performance}

In general, the performance of any detection method can be assessed through comparing different outcomes; true positives ($TP$), false positives ($FP$), false negatives ($FN$), or true negatives ($TN$). A $TP$ occurs when a real source is correctly identified, while an $FP$ arises from a spurious detection. Conversely, an $FN$ corresponds to a missed source, and a $TN$ indicates correctly identifying empty background. These basic outcomes can be combined to calculate performance metrics \citep[e.g.,][]{cormack_statistical_2006, manning_introduction_2008, beitzel_map_2009, sortino_radio_2023}. Below, we summarise the metrics used to evaluate each source-finder:

\begin{equation*}
    \text{Recall} = \frac{TP}{TP + FN}~\,\text{(or Completeness)}~,
\end{equation*}

\begin{equation*}
    \text{Precision} = \frac{TP}{TP+FP}~\,\text{(or Reliability)}~,
\end{equation*}

\begin{equation*}
    \text{Informedness} = \frac{TP}{TP+FN} - \frac{FP}{TN+FP}~.
\end{equation*}

\noindent Here recall is the fraction of the truly positive instances, precision is the proportion of predicted positive instances that are actually correct. Informedness provides a single measure of the detection methods performance relative to random chance. The precision-recall curve, $P(R)$, describes the relationship between precision and recall as some threshold is varied \citep{beitzel_map_2009, gupta_deep_2023}. A common evaluation metric, average precision (AP), is determined through integrating the precision-recall curve at a fixed intersection-over-union (IoU) threshold \citep{manning_introduction_2008}. The mean average precision (mAP) is then the average of AP values computed across multiple IoU thresholds \citep{cormack_statistical_2006, beitzel_map_2009}:

\begin{align*}
    AP &= \int_0^1 P(R)dR~, &mAP = \frac{1}{n} \sum_{n} AP_{n}~.
\end{align*}

Each of the three source-finders considered in this work were evaluated using slightly different sets of performance statistics, as reported in their respective papers. \citet{Gordon_DRAGN_2023} report the performance of \textit{DH} in terms of completeness and reliability. They estimate the completeness to be $\geq 45\%$ for sources with $S_{3,\text{GHz}} > 20\,$mJy, rising to $\geq 85\%$ for $S_{3,\text{GHz}} > 100\,$mJy. The reliability of their catalogue is reported as $89^{+1.2}_{-1.6}\%$.
\citet{Segal_Complex_2019, segal_complexity_2023} determine the performance in a slightly different way, using recall and informedness (reported as $90\%$ and $84\%$, respectively). \citet{Gupta_rgcat_2024} instead make use of the average precision measurement to quantify the performance of RG-CAT, with an $AP_{50}$ (average precision at IoU = 0.5) of $73.2\%$ for the radio galaxy bounding box predictions and $71.7\%$ for the infrared host keypoint positions.

In \S\ref{rgcat_sec}, we showed the performance of RG-CAT for EMU-G09 to be similar to that for EMU-PS in \citet{Gupta_rgcat_2024}. The performance of CG-Complexity is likely to be very similar to the quoted values from \citet{Segal_Complex_2019}, as it was evaluated on radio data with comparable observational parameters. For \textit{DH} however, due to the better sensitivity and poorer resolution of EMU relative to VLASS, we cannot assume that the performance metrics will be the same. It is likely that the \textit{DH} algorithm will produce somewhat more spurious doubles compared to the VLASS proportions, mainly due to the greater sensitivity (i.e., more nearest neighbour pairs). Formally determining the completeness and reliability of \textit{DH} on EMU data would require using either simulated sources or injections into real images. Such analysis is beyond the scope of this work, however we performed a visual inspection of 1,000 randomly selected \textit{DH} detections ($\sim\!37\%$ of \textit{DH} sources, assessed in ten independent subsets of 100 by L.J.B) to obtain a lower limit on the reliability. Approximately $52.5\%$ of sources appear to correspond to genuine double-lobed radio galaxies. This reliability is likely underestimated, as visual inspection alone, particularly at EMU’s sensitivity and resolution, cannot always confirm whether a faint pair of components originates from the same physical system. Excluding sources with component flux ratios $<10$ or low S/N would likely further increase this reliability estimate.

While these performance statistics indicate that each source-finder performs reliably within its own design framework, the differences in depth, resolution, and evaluation criteria, mean that we cannot compare these metrics directly. Through applying each source-finder to the same field, we can instead assess the sensitivities and biases of each approach, and explore the types of radio sources for which each source finder is most effective at identifying.

\subsection{Sources detected in G09}\label{sources_in_g09}

\begin{figure*}[htp!]
    \centering
    \includegraphics[width=\linewidth]{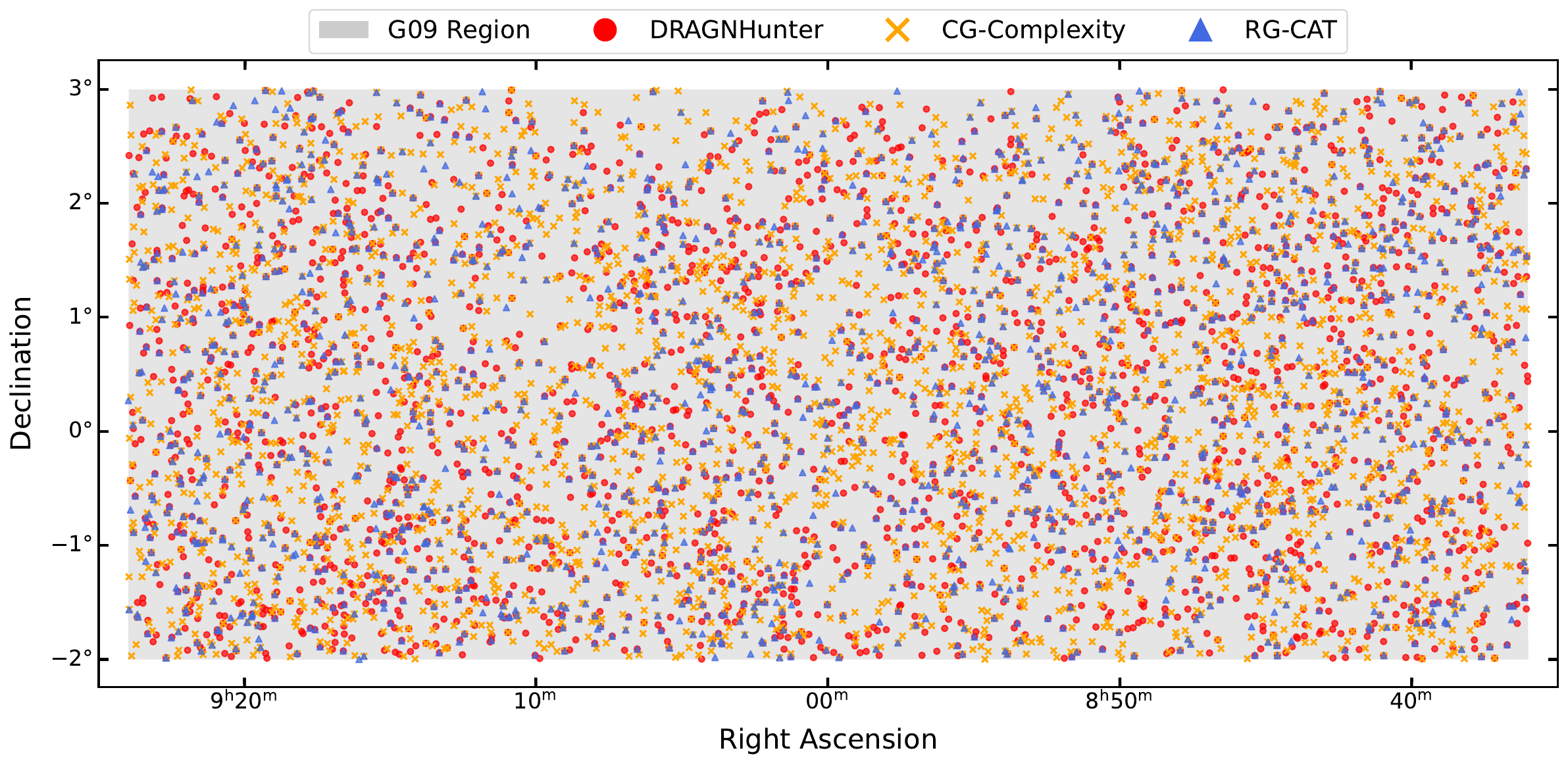}
    \caption{EMU field (grey), overlaid with the sky positions of sources detected by each finder: likely DRAGNs identified by DRAGN{\scriptsize HUNTER} (red dots), significantly complex regions from CG-Complexity (orange crosses), and extended sources detected by RG-CAT (blue triangles). At this scale, it is evident there is clustering of sources in some regions, and others showing a lack of detections from all finders. There is also very little overlap between each source-finder.}
    \label{g09_all_sources}
\end{figure*}

We show the positions of the likely DRAGNs (\textit{DH}), highly complex regions (CG-complexity) and extended sources (RG-CAT) detected in the G09 region in Figure~\ref{g09_all_sources}. The distributions of sources are not uniform across the field. Clustering on square degree scales is evident in certain regions, while other areas show a lack of detections from all finders. This spatial variation reflects underlying large-scale  structure \citep[][]{Driver_GAMA_2011}, and the filaments and voids in the radio source population should be investigated in future work. Table~\ref{source_counts} summarises the total number of sources detected by each finder, with the corresponding source density calculated for EMU-G09. The higher number of detections from CG-complexity likely reflects its sensitivity to a wide range of morphologies, including irregular or multi-component emission that may not meet the criteria of \textit{DH} and RG-CAT (as well as the detected point sources). For adjacent pairs of well-resolved sources, CG-Complexity may also identify each lobe of a radio galaxy as a significantly complex source.

\begin{table}[h!]
\centering
\caption{Number of sources detected in EMU-G09 field by each extended-source finder, and the corresponding surface densities.}
% \resizebox{\linewidth}{!}{%
\begin{tabular}{lcc}
\toprule
Approach & Sources & Sources/deg$^{2}$ \\
\midrule
DRAGN{\footnotesize HUNTER} & $\numdragnsG$ & $\dragndensityG$ \\
CG-Complexity & $\numcompsG$ & $\compdensityG$ \\
RG-CAT & $\numrgcatG$ & $\rgcatdensityG$ \\
\bottomrule
\end{tabular}%
% }
\label{source_counts}
\end{table}

\begin{figure}[h!]
    \centering
    \includegraphics[width=\linewidth]{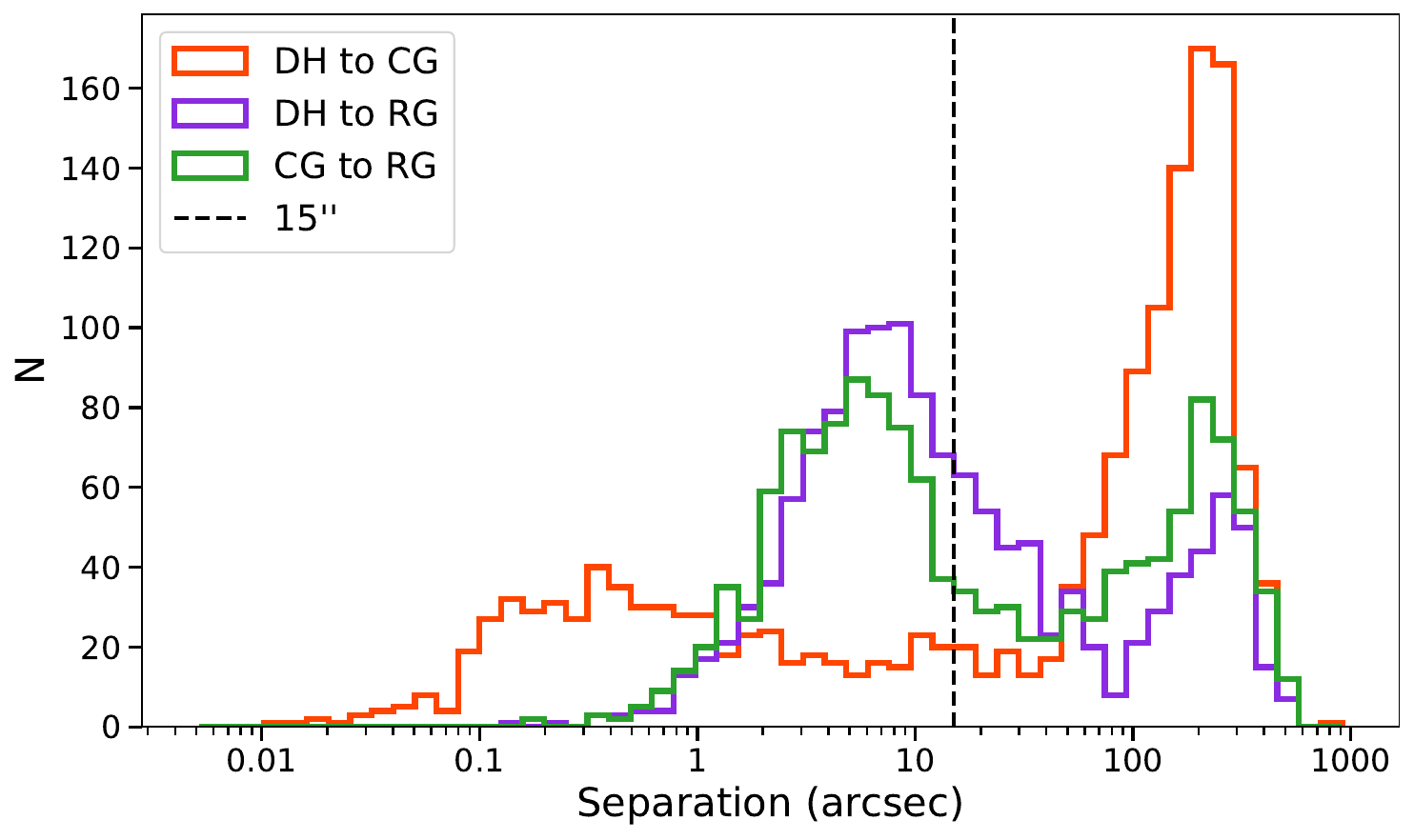}
    \caption{Separation distributions for matches between source-finder catalogues: DH-CG (dark orange), DH-RG (purple), and CG-RG (green). Each distribution is bimodal, where the first peak is likely due to genuine matches and the second is likely dominated by random associations. We adopt a $15''$ radius as the threshold for genuine matches.}
    \label{det_overlap_rad}
\end{figure}

To determine a reasonable matching radius for comparing sources detected by each finder, we first compute the number of matches for each unique pair of our EMU-G09 source-finder catalogues out to $1^\circ$: DRAGN{\footnotesize HUNTER} to CG-Complexity (DH-CG), DRAGN{\footnotesize HUNTER} to RG-CAT (DH–RG), and CG-Complexity to RG-CAT (CG-RG), with $1605$, $1349$, and $1362$ matches, respectively. Figure~\ref{det_overlap_rad} shows the distribution of separations for these matched pairs. In each case, the distribution is bimodal: the first peak at small separations likely corresponds to genuine matches, while the second peak arises from random associations. For all catalogue pairs, the first peak diminishes around $15''$, above which random associations dominate. We therefore adopt a matching radius of $15''$ to define overlap between sources identified by each of our finders.

\begin{figure}[h!]
    \centering
    \includegraphics[width=0.9\linewidth]{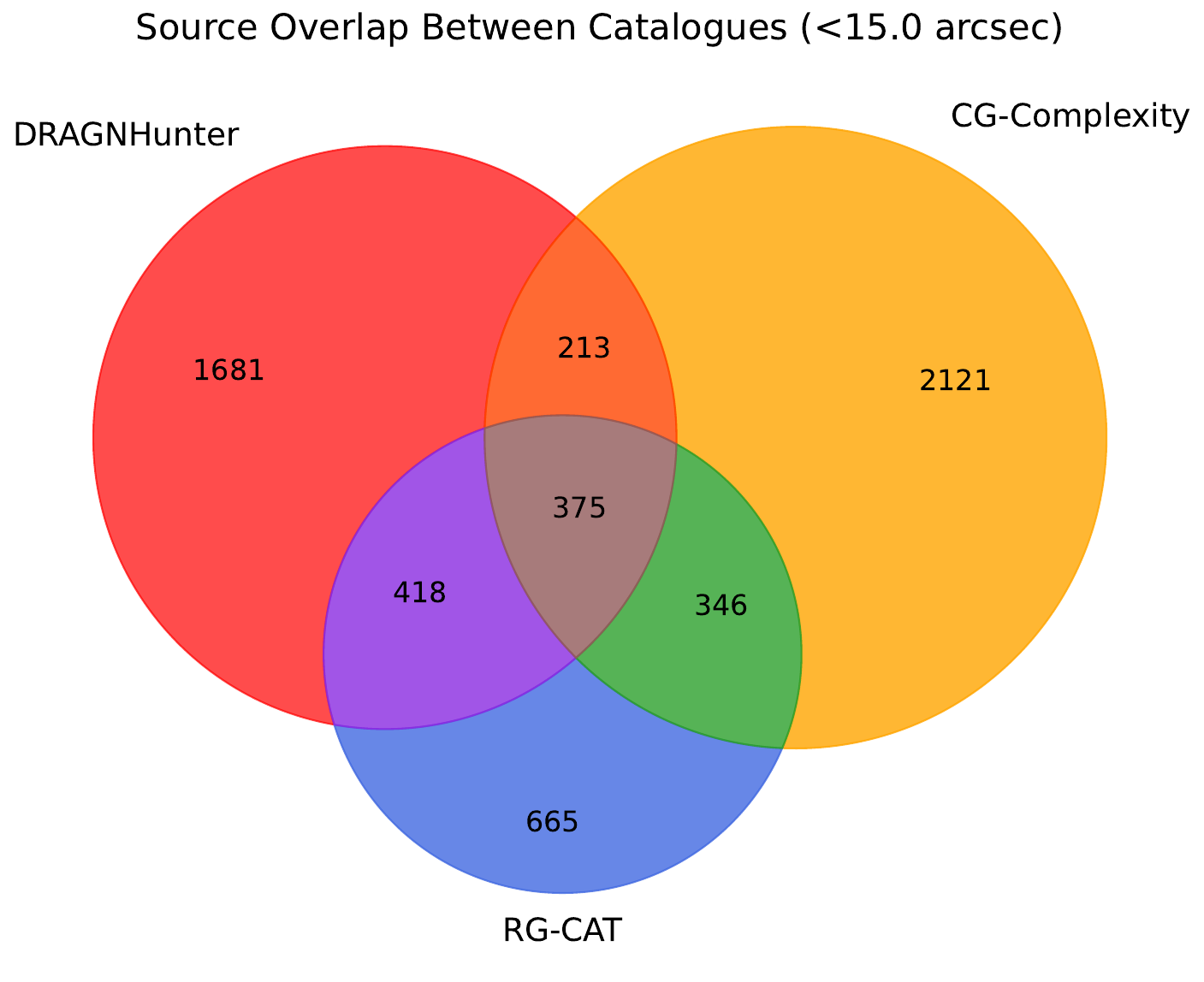}
    \caption{Overlap ($<15''$) of sources detected by DRAGN{\scriptsize HUNTER} (red), CG-Complexity (orange), and RG-CAT (blue). Each region of the diagram represents the number of sources uniquely or jointly identified by the corresponding source-finders. Only $375$ sources are common to all three, with this small overlap likely reflecting the differing selection biases of each method.}
    \label{pos_overlap}
\end{figure}

Detections from two or more source-finders within $15''$ are likely to belong to the same source. Figure~\ref{pos_overlap} establishes that there is limited overlap between each approach with $375$ sources being independently detected by all source-finders. This is likely due to the different sources that each source-finder is tailored to find. As the primary selection criteria for \textit{DH} is a nearest neighbour, it may be biased towards smaller sources (i.e., smaller angular separation between components, Figure~\ref{DRAGNHunter_LAS_Flux}). RG-CAT, due to being trained on manually identified sources, may be biased towards sources that are larger and brighter (i.e., sources more easily identified by-eye). The CG-complexity measure is agnostic to predetermined morphology types, and is instead searching for significantly complex regions. These regions would often coincide with an extended source such as a DRAGN, but also includes irregular diffuse structure. As CG-complexity is applied to the cutouts of the \textit{Selavy} islands, this can possibly explain why there are more detections than \textit{DH} and RG-CAT (i.e., each lobe of a  large, well-resolved radio source may be deemed significantly complex).

\begin{figure}[h!]
    \centering
    \begin{subfigure}{\linewidth}
       \centering
       \includegraphics[width=\linewidth]{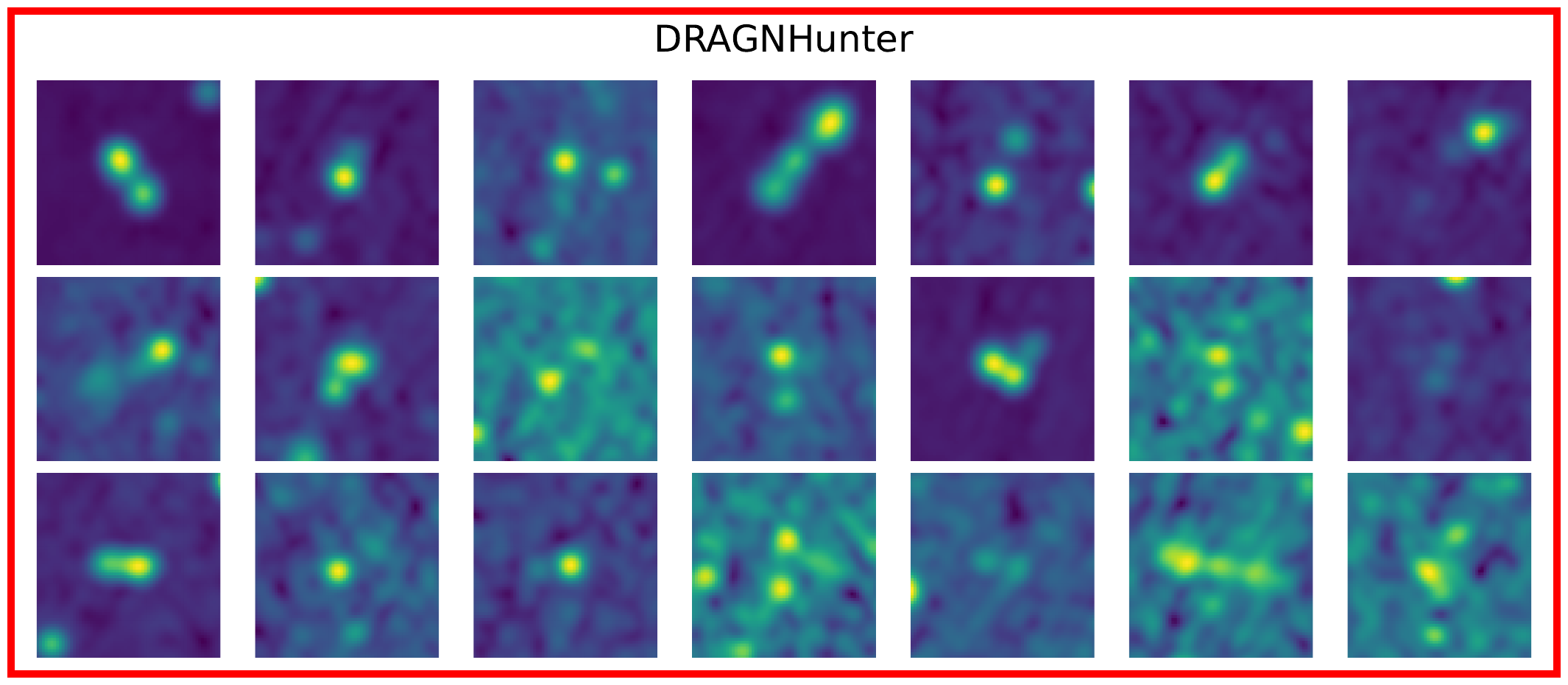}
    \end{subfigure}
    \begin{subfigure}{\linewidth}
        \centering
        \includegraphics[width=\linewidth]{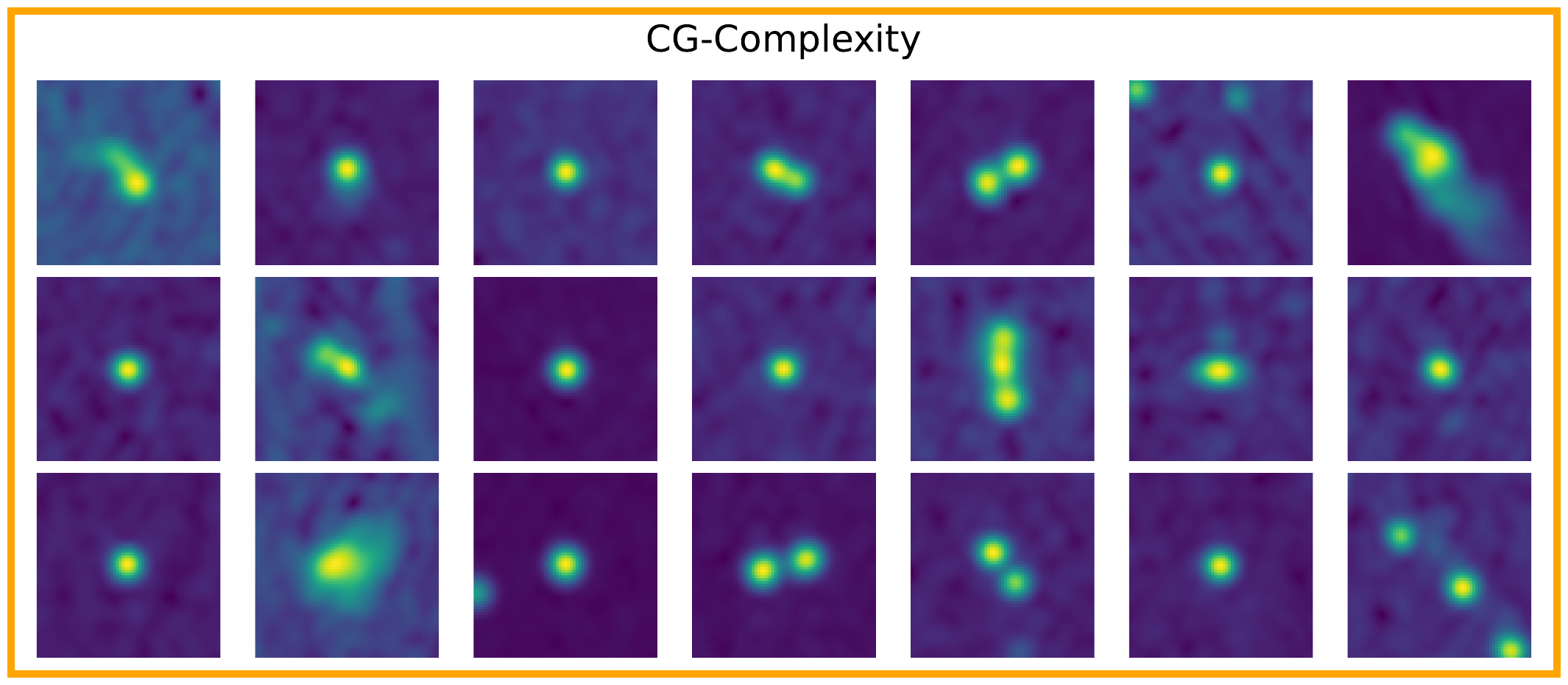}
    \end{subfigure}
    \begin{subfigure}{\linewidth}
        \centering
        \includegraphics[width=\linewidth]{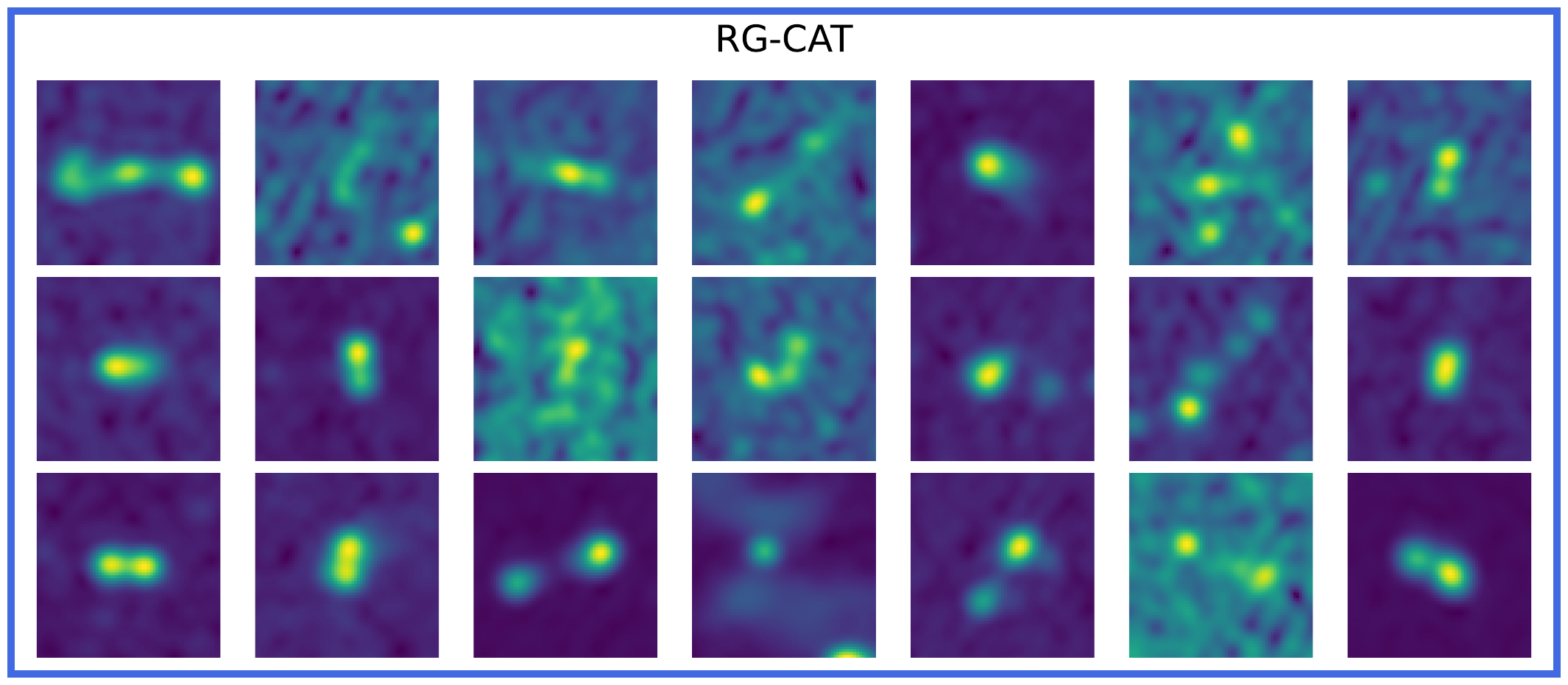}
    \end{subfigure}
    \caption{Example sources detected by each source-finding approach. Each panel contains a random selection of 21 sources identified by each of the respective finders. All cutouts are $64\times64$ pixels, corresponding to an angular size of $128''\times128''$. \textit{DH} appears to predominantly identify typical double-lobed structure; CG-Complexity identifies compact sources, typical doubles and more diffuse irregular structure; RG-CAT tends to identify typical doubles as well as more complex extended structure.}
    \label{eg_sources}
\end{figure}

To verify this, we show cutouts of the CG-Complexity sources, as well as sources from \textit{DH} and RG-CAT in Figure~\ref{eg_sources}. The \textit{DH} sample is dominated by classical double-lobed radio galaxies, with \textit{DH} sources generally appearing to be relatively compact. This is reflective of the nearest-neighbour pairing strategy used by \textit{DH}, which favours close, symmetric component pairs. Many \textit{DH} cutouts also include faint, low S/N sources where Gaussian components have been fitted to marginal detections or possibly to bright noise fluctuations. Conversely, the RG-CAT sources often appear to be slightly larger and brighter than the \textit{DH} sources. CG-Complexity identifies typical radio doubles as well as regions of more intricate or irregular emission. Many cutouts in the random selection of CG-Complexity sources show a simple compact source where, as discussed in \S\ref{cg_comp_section}, the surrounding structure can increase the overall complexity score.

\begin{figure}[h!]
    \centering
    \includegraphics[width=\linewidth]{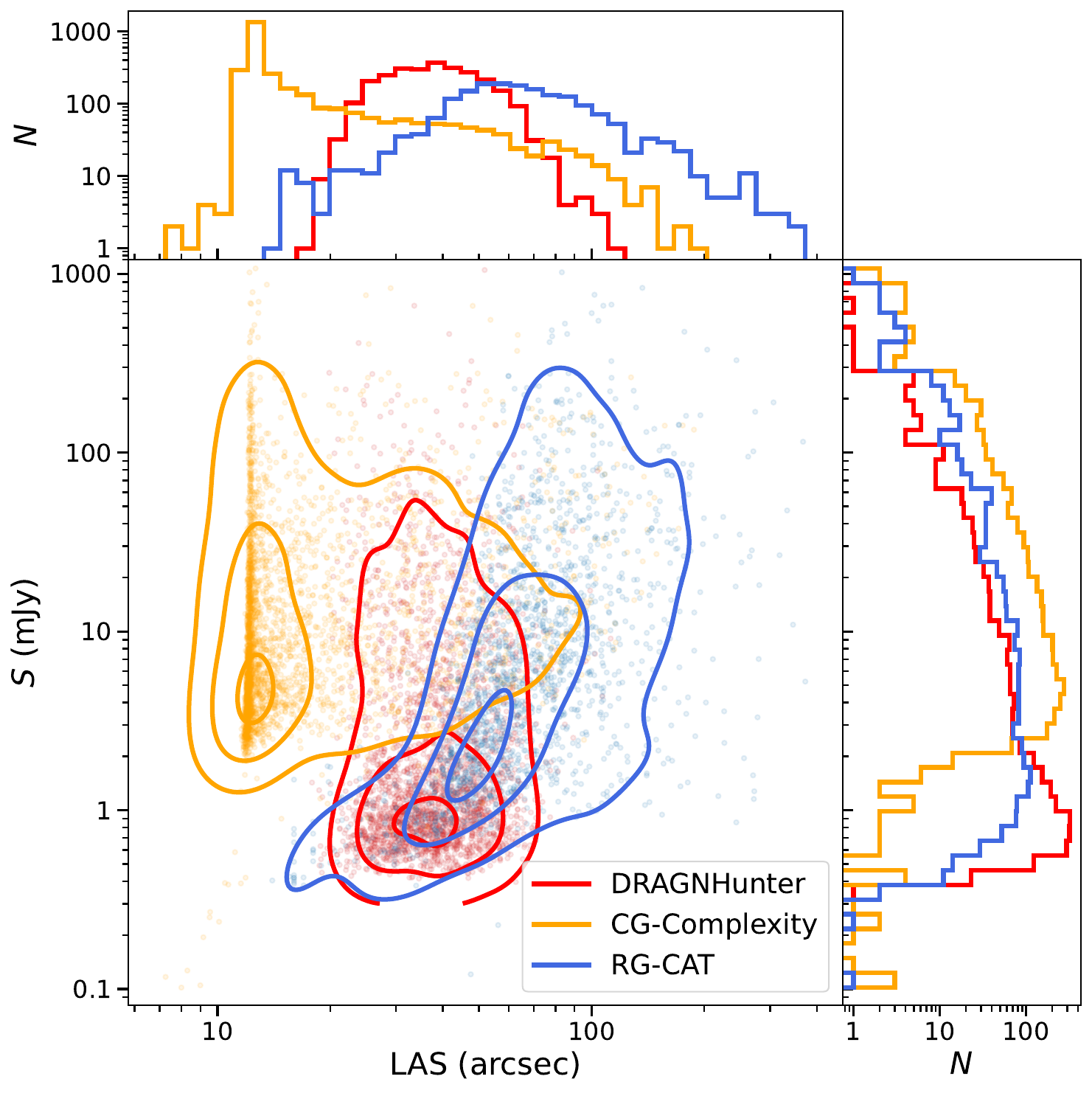}
    \caption{Distributions of LAS and integrated flux density (S) for sources detected by each source-finding approach. Density contours for the distributions of LAS and $S$ for DRAGN{\scriptsize HUNTER} sources are shown in red, high complexity islands in orange, and RG-CAT sources in blue. Contours contain $90\%$, $50\%$, and $10\%$ of the data points. While there is substantial overlap in this parameter space, CG-Complexity includes a population of smaller-angular-size sources, whereas RG-CAT tends to detect larger and brighter systems, again likely reflecting selection biases inherent to each method.}
    \label{all_las_flux}
\end{figure}

To further understand the populations of sources identified by each source-finder, we compare the largest angular size (LAS), and flux density ($S$) in Figure~\ref{all_las_flux} for all sources. Sources identified by all finders generally exhibit similar flux densities indicating that flux is not the primary driver of detection differences. While there is significant overlap in this parameter space, there are some obvious differences, particularly in LAS, between sources in our sample. The CG-Complexity sources tend to have the smallest angular size when compared to \textit{DH} and RG-CAT sources, with RG-CAT being the largest. This is likely due to how LAS is defined for each source finder. For \textit{DH} sources, LAS is the separation between the component positions, increased slightly by the semi-major size of each component. For the RG-CAT sources, LAS corresponds to the major axis of a bounding box predicted by Gal-DINO, designed to encapsulate the entire source emission. It is therefore likely that LAS of RG-CAT may be slightly overestimated for a given source, and slightly underestimated for \textit{DH} sources. This is also expected given that RG-CAT is trained on visually identified radio galaxies, where bounding boxes are drawn to include all detectable emission, naturally favouring larger angular extents. For the CG-Complexity sources, LAS is the size of the major axis of each \textit{Selavy} island, determined by the number of pixels brighter than minimum flux density threshold. As an island may correspond to a radio galaxy core, or a single lobe it is expected to have a smaller LAS than an entire extended radio source.

\begin{figure}[h!]
    \centering
    \includegraphics[width=0.95\linewidth]{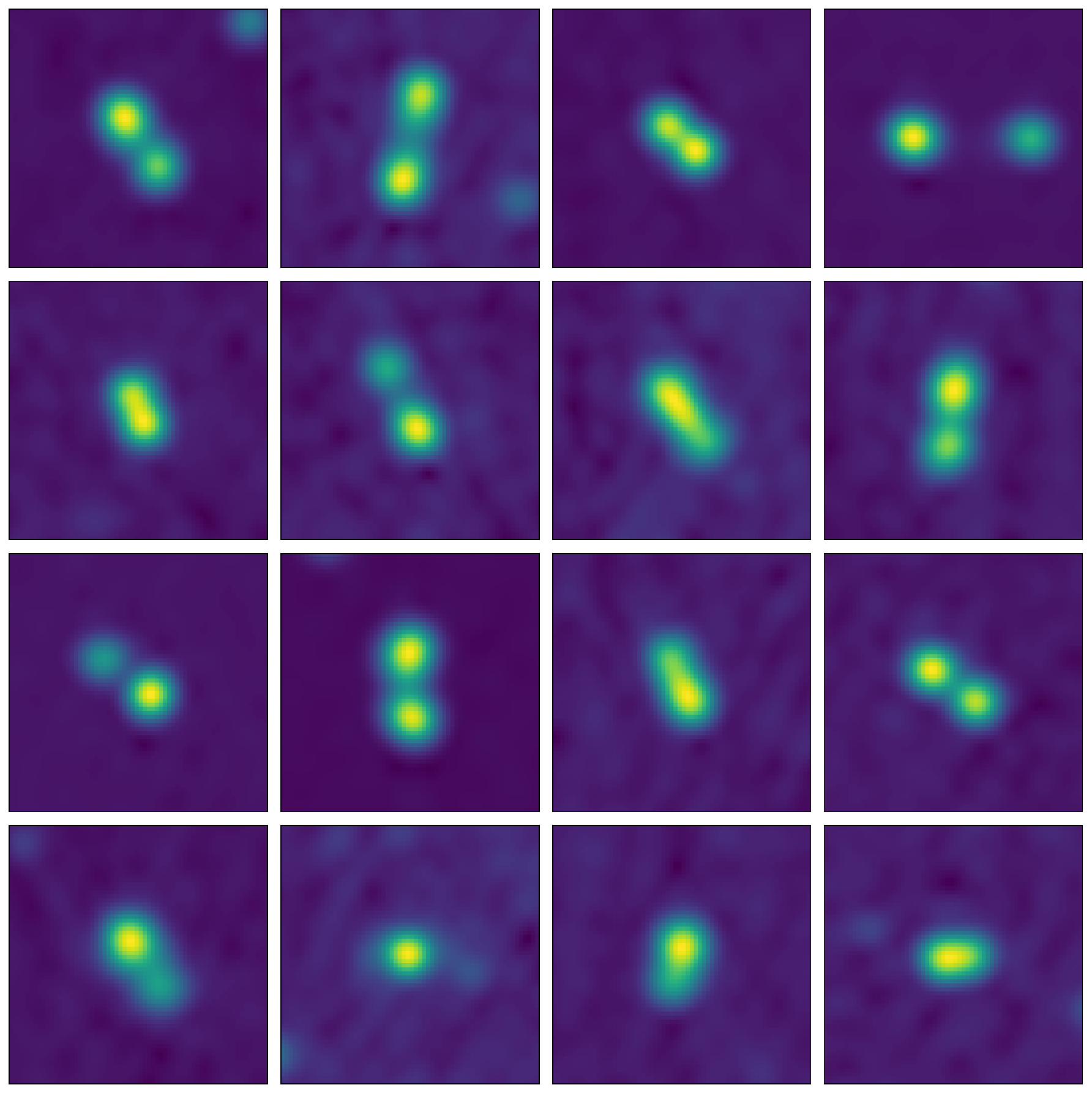}
    \caption{A random selection of $16$ sources detected by all source-finders. These objects typically exhibit well-defined double-lobed morphologies of moderate angular size. All cutouts are $64\times64$ pixels, corresponding to an angular size of $128''\times128''$.}
    \label{common_sources}
\end{figure}

In Figure~\ref{common_sources} we highlight a representative sample of sources that were detected by all finders. These appear to be classical double-lobed radio galaxies, with relatively bright (average of $\sim\!31$~mJy), moderately extended emission (average of $\sim\!51''$). Their morphology places them in the narrow regime where the strengths of the three methods overlap: distinct nearby components for \textit{DH}, sufficient structural complexity to be identified by CG-Complexity, and structures similar to observed FR-I/FR-II type sources included in RG-CAT’s training set. As a result, they form the most stable and consistently recovered population across the different detection strategies.

However, due to the limited positional overlap between sources identified by each finder (Figure~\ref{pos_overlap}), the systematically smaller LAS values measured for CG-Complexity sources (compared to \textit{DH} and RG-CAT) likely reflect the presence of two distinct populations within the G09 field: a relatively compact population dominated by isolated components or cores, and a more extended population of complex radio sources. This interpretation is consistent with results from the GLEAM 4-Jy sample \citep[G4Jy;][]{sarah_g4JyI_2020, sarah_g4JyII_2020, sarah_salt_2025}, where the size distribution similarly indicates a more compact, marginally resolved population, and an extended radio source population. The marginally resolved population may also correspond to sources at higher redshift.

\begin{figure}[h!]
    \centering
    \includegraphics[width=\linewidth]{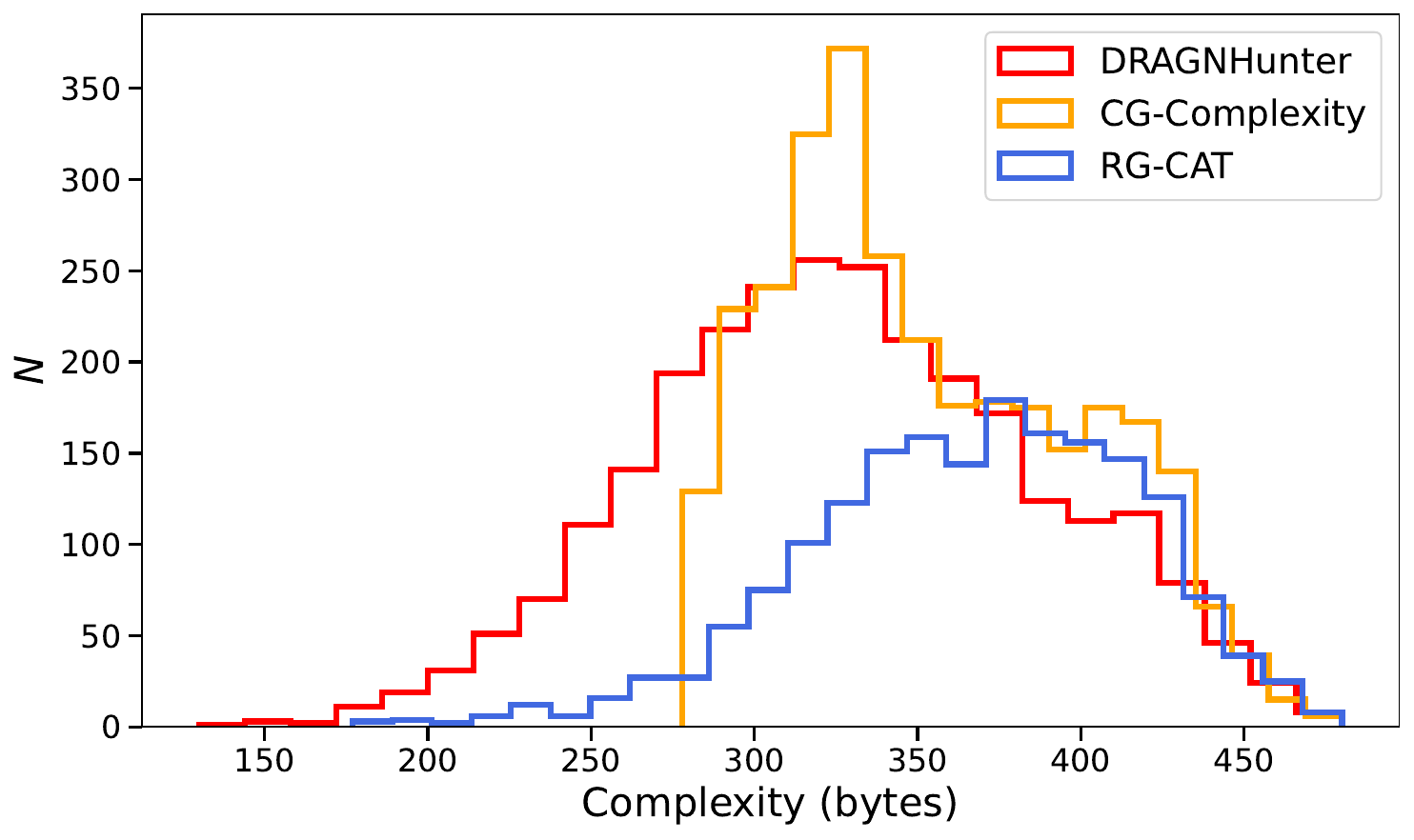}
    \caption{Comparison of calculated complexity of sources detected by DRAGN{\scriptsize HUNTER} (red), CG-Complexity (orange), and RG-CAT (blue). All sources identified by each approach have comparable apparent complexity, with RG-CAT tending to pick up the higher complexity sources. This likely reflects its training on visually classified, often larger and more irregular sources, while \textit{DH} tends to favour more symmetric, double-lobed morphologies.}
    \label{complexity_comparison}
\end{figure}

We also determined the CG-Complexity of the \textit{DH} and RG-CAT sources by generating $64\times64$ pixel cutouts centred on their positions, applying the same processing steps as for the \textit{Selavy} islands. The resulting distributions are shown in Figure~\ref{complexity_comparison}. Sources identified by all approaches exhibit broadly comparable apparent complexity, with RG-CAT tending to detect slightly more complex systems than \textit{DH}. This difference likely reflects the design of the two methods: \textit{DH} primarily targets classical double-lobed radio galaxies, which can often appear as two relatively symmetric and circular lobes, and thus show comparatively lower apparent complexity. In contrast, RG-CAT, trained on visually identified DRAGNs, may be biased toward larger and brighter sources that naturally exhibit more extended or irregular structure. Further, the inclusion of morphological tags within RG-CAT, specifically the `peculiar' class, is reserved for atypical irregular morphologies, therefore allowing it to capture a broader range of complex radio emission than \textit{DH}.

Our results show that each method captures a different subset of the extended radio source population in the G09 field. \textit{DH} appears to be effective at identifying classical double-lobed systems, while RG-CAT recovers slightly larger and morphologically complex objects. CG-Complexity in turn, identifies diffuse or irregular emission that may not follow traditional radio-galaxy morphologies. Although the three approaches together detect nearly all sources in the field, none provides a complete census of the extended-source population on its own. Certain classes, such as giant radio galaxies \citep[GRGs;][]{dabhade_discovery_2017, dabhade_grg_2020, andernach_grg_25}, remain difficult to capture without additional or visual inspection. Tailoring algorithms to such rare morphologies could improve completeness for specific classes but would risk biasing detection and increasing false positives in other regimes.

\subsection{Multiwavelength Cross-matching}

\subsubsection{WISE}

We make use of the Wide-field Infrared Survey Explorer mission \citep[WISE,][]{wright_wise_2010}, which provides coverage for the entire sky in four mid-infrared bands, $3.4$, $4.6$, $12$, and $22 \,\mu$m ($\rm{W}1\, \rm{to}\,\rm{W}4$, respectively). Infrared cross-matching is implemented in a slightly different way for each source-finder presented in this work. The \textit{DH} script incorporates cross-matching to the AllWISE catalogue \citep[][]{cutri_allwise_2012, cutri_allwise_2013} for candidate hosts of identified DRAGNs (if one exists), or to the flux weighted centroid position between lobes (for sources without a candidate host). This cross-matching is done by querying the AllWISE database for sources in a $30''$ search radius and selecting the most likely AllWISE counterpart via likelihood ratio testing of sources within the search radius. In our sample, the \textit{DH} script identified $1748$ ($65.1\%$) sources with an AllWISE counterpart.

For CG-Complexity sources, we perform a positional cross-match with the AllWISE catalogue using a search radius of $5''$ to identify potential nearby hosts. Since the positions of some islands may correspond to single radio lobes, offset from the actual host galaxy, we expect some of these matches to be spurious. However nearby, large, and well resolved sources are much more rare than smaller more compact sources, so we do not expect many single-lobe islands in our sample. We find $1900$ ($62.2\%$) AllWISE counterparts for the highly complex islands. The RG-CAT pipeline incorporates cross-matching with the CatWISE catalogue \citep[][]{Marocco_catwise_2021} for identified radio sources, however the CatWISE catalogue only contains $\rm{W}1$ and $\rm{W}2$ magnitude measurements. Following \citet{Gupta_rgcat_2024}, we cross-match the CatWISE positions to the AllWISE catalogue within $1''$. Not all sources are detected in CatWISE however, and some CatWISE sources may not have an AllWISE match (either because they are too faint for AllWISE, or because AllWISE is a blend of two or more CatWISE sources). If the CatWISE to AllWISE search returns no matches, we then use the radio coordinates with a search radius of $5''$.  Using this methodology, we find $922$ ($50.6\%$) AllWISE counterparts for the extended RG-CAT sources,  with $644$ from the CatWISE positions and $278$ from the radio positions. 

The \textit{DH} script includes a likelihood ratio (LR) calculation that provides both an LR and a corresponding reliability estimate for each cross-matched source (see \ref{DH_LR} for further details). Using the derived reliability values, we find the false-positive rate (fpr) for \textit{DH} sources cross-matched with AllWISE to be approximately $\sim\!10.2\%$. For CG-Complexity and RG-CAT sources, we estimate the fpr by artificially shifting both the RA and Dec values of the AllWISE catalogue by $1^{\circ}$ and repeat the cross-match as described above. This resulted in $282$ and $182$ matches for the CG-Complexity and RG-CAT samples, corresponding to an approximate fpr of $\sim\!9.2\%$ and $\sim\!10.0\%$, respectively. 

\subsubsection{GAMA}

\begin{figure}[h!]
    \centering
    \includegraphics[width=\linewidth]{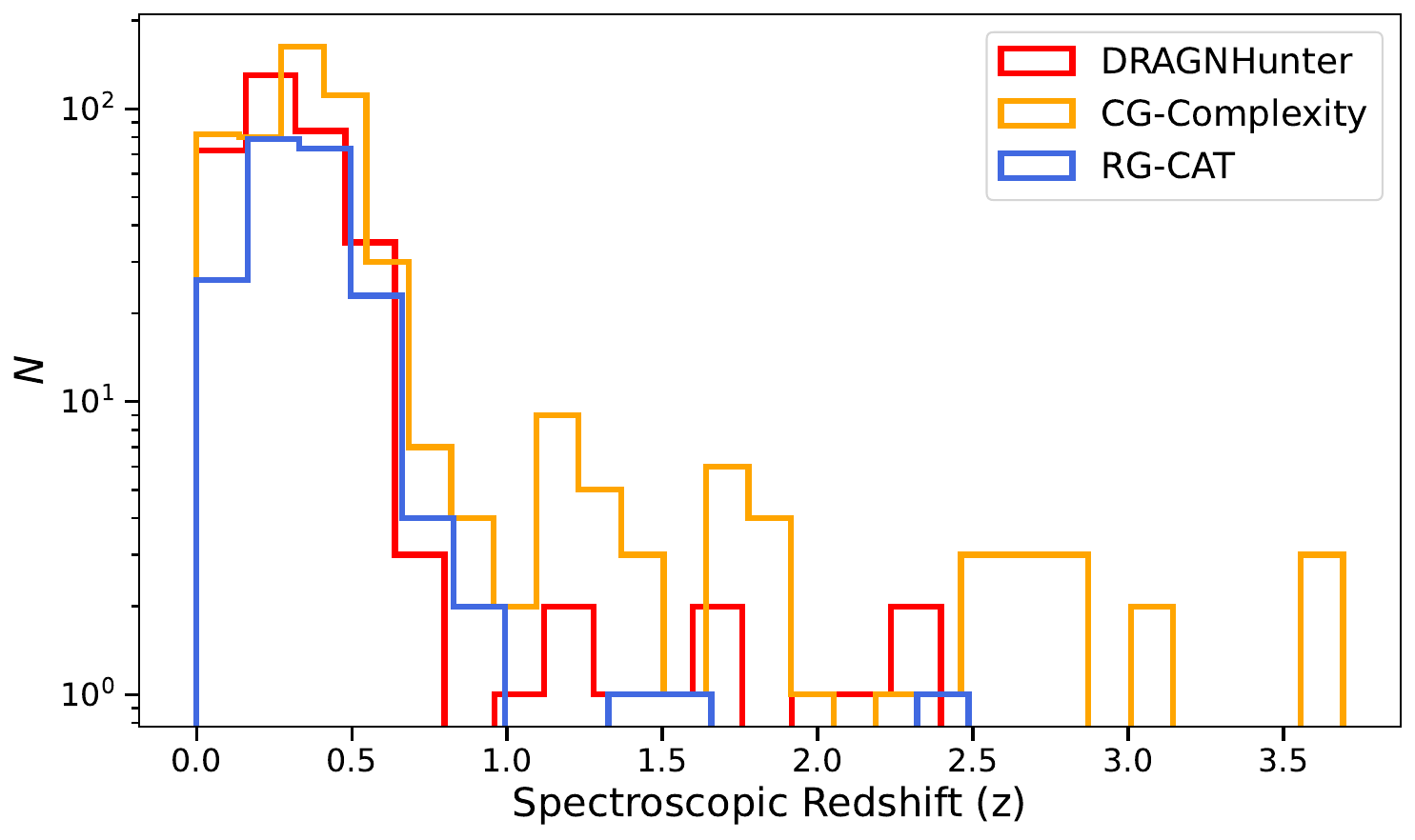}
    \caption{Spectroscopic redshift distributions of \textit{DH}, CG-Complexity and RG-CAT sources cross-matched with GAMA-DR4 data (red, orange, and blue solid lines, respectively). All redshifts here have a quality factor $>2$. The \textit{DH} and RG-CAT sources exhibit similar redshift distributions, while CG-Complexity extends to systematically higher redshifts, consistent with its sensitivity to smaller or more compact emission regions that may correspond to unresolved high-redshift doubles.}
    \label{sf_z_comparison}
\end{figure}

We then match the AllWISE coordinates associated with the \textit{DH}, CG-Complexity, and RG-CAT sources with our GAMA catalogue (\S\ref{gama_section}) using a $2''$ search radius. This resulted in $675$, $690$, and $355$ matches, for \textit{DH}, CG-Complexity and RG-CAT, respectively. The distributions of spectroscopic redshifts (spec-z) from GAMA are shown in Figure~\ref{sf_z_comparison}. Only spec-z values with a redshift quality $>2$ are shown ($335$, $524$, and $210$ sources for \textit{DH}, CG-Complexity, and RG-CAT, respectively). For subsequent analyses requiring redshift, we only use sources with a quality factor$>2$. The redshift distributions of the \textit{DH} and RG-CAT sources are closely aligned, whereas CG-Complexity sources are found to extend to systematically higher redshift. This difference likely reflects a combination of selection and methodological effects. While some contribution may arise from spurious or chance associations in the CG sample, this alone cannot account for the observed shift. As CG-Complexity operates directly on \textit{Selavy} islands, high-redshift doubles that are unresolved at EMU resolution may be represented as a single, moderately extended component. Such structures can still exhibit non-uniform or irregular brightness distributions that register as ``complex’’ by the CG-Complexity metric, even when not recognised as multi-component sources by algorithms like \textit{DH} or RG-CAT.

\section{Astrophysical Properties of Sources in G09}

\begin{figure}[h!]
    \centering
    \includegraphics[width=\linewidth]{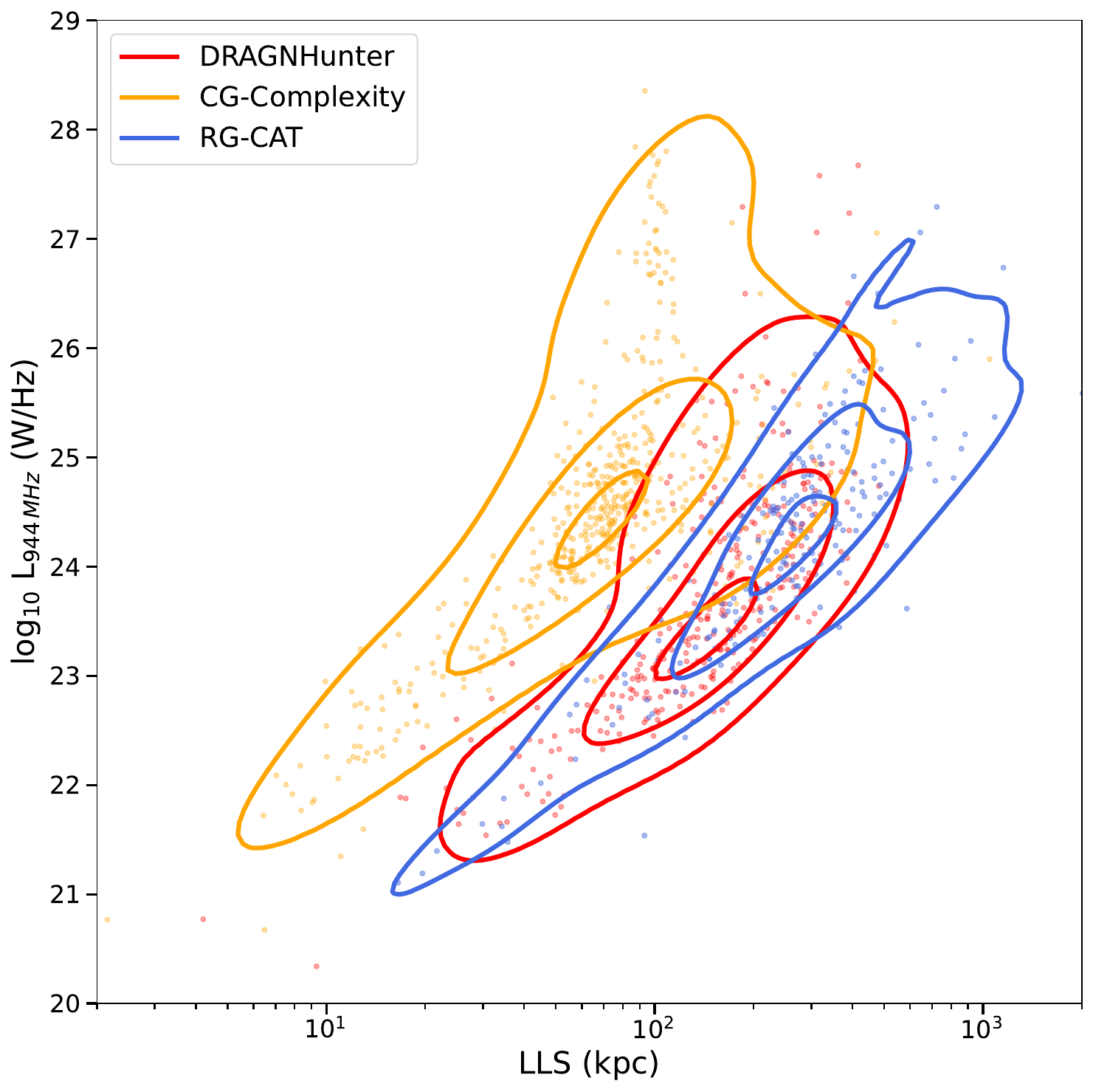}
    \caption{Distributions of LLS and radio luminosity for sources identified by \textit{DH} (red), CG-Complexity (orange), and RG-CAT (blue). Contours enclose $90\%$, $50\%$, and $10\%$ of the data points. LLS values are calculated using LAS, which is defined differently for each finder (see \S~\ref{sources_in_g09})}
    \label{sf_lum_lls}
\end{figure}

We determine both the luminosity ($L_{944\,\rm{MHz}}$) and the largest linear size (LLS) for our sample of sources (Figure~\ref{sf_lum_lls}). There is significant overlap between the RG-CAT and \textit{DH} sources, however as before, the RG-CAT sources tend to be slightly larger than \textit{DH} sources. The high complexity sources, conversely, always tend to be smaller than both \textit{DH} and RG-CAT sources. Particularly for the \textit{DH} and CG-Complexity sources, we see there is a steep increase in the luminosity of the sources at $L_{944\,\rm{MHz}}\approx\!10^{24}\,\rm{W/Hz}$, which may be indicative of two populations (i.e., FR-I/FR-II, for example).

\begin{figure}[h!]
    \centering
    \includegraphics[width=\linewidth]{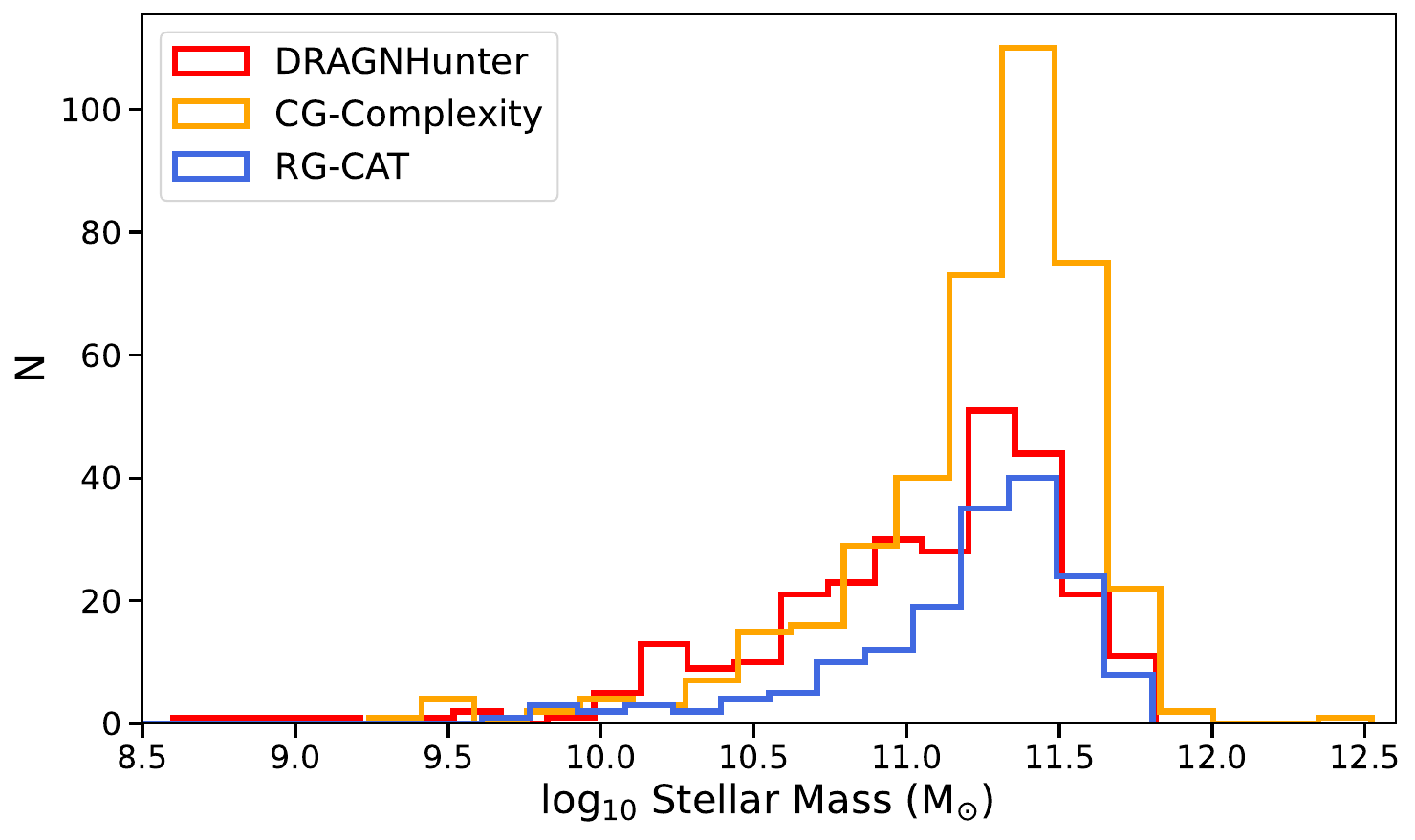}
    \caption{Stellar mass distributions of \textit{DH}, CG-Complexity and RG-CAT sources cross-matched with GAMA-DR4 data (red, orange, and blue solid lines, respectively). All three distributions are broadly similar, peaking around $10^{11.4}\,M_\odot$.}
    \label{sf_masses}
\end{figure}

We adopt the stellar masses estimated by \citet{taylor_stellar_2011} determined using a Chabrier initial mass function (IMF), of the host galaxies identified for each source finder. We show the mass distributions in Figure~\ref{sf_masses}. All three distributions are broadly similar, with each peaking around $10^{11.4}\,M_\odot$, indicating that the typical host of an extended radio source in the G09 field is a massive galaxy. \textit{DH} tends to identify more low mass sources than RG-CAT, with a number of low-mass hosts comparable to those from CG-Complexity. Despite the largely distinct source samples identified by each finder, this similarity in stellar mass suggests that the underlying galaxy populations are comparable in terms of their fundamental properties, and that differences in source detection are primarily driven by the morphological or structural criteria of the source-finders rather than by host galaxy characteristics. 

We implement the WISE colour diagnostic \citep[$\rm{W}1-\rm{W}2$ versus $\rm{W}2-\rm{W}3$ magnitudes,][]{lake_wisediagnostic_2012} to classify the infrared (IR) properties of the host galaxies for sources identified by each respective source-finder. Here we adopt the limits defined by \citet{mingo_agndiagnostic_2016},  where the $\rm{W}1-\rm{W}2$ colour separates AGN dominated systems ($> 0.5$) from those dominated by host-galaxy light ($< 0.5$). For host-dominated sources, the $\rm{W}2-\rm{W}3$ colour further distinguishes between passive/elliptical galaxies ($<1.6$), star-forming galaxies ($1.6 - 3.4$), and starburst systems ($>3.4$), which often correspond with ultra-luminous infrared galaxies (ULIRG). We only apply these classifications to sources with S/N in all three bands $>3$. While this requirement biases the distribution against passive galaxies, whose $\rm{W}3$ emission is often several magnitudes fainter than $\rm{W}1$ or $\rm{W}2$ \citep[e.g.,][]{cluver_powerofwise_2020}, the higher S/N threshold means that the classifications we do obtain are relatively robust. Therefore, the resulting colour distribution should be interpreted as fairly reliable, but not representative of the entire sample.

\begin{figure}[h!]
    \centering
    \includegraphics[width=\linewidth]{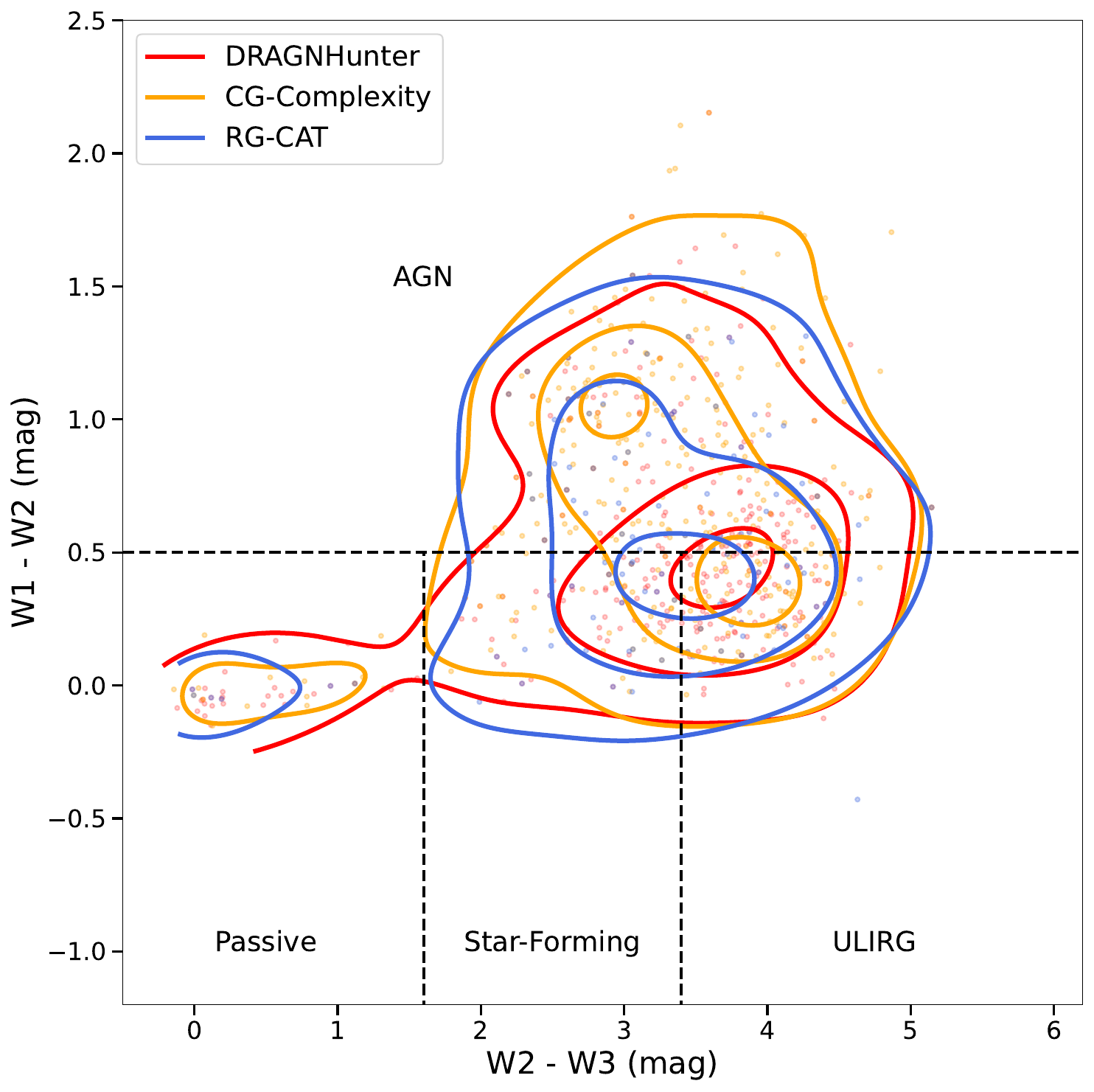}
    \caption{The WISE colour distributions for the hosts of sources identified by \textit{DH} (red), CG-Complexity (orange), and RG-CAT (blue). The contours contain $90\%$, $50\%$, and $10\%$ of the data points. Despite differences in detection strategy, sources identified by all three approaches have similar infrared host populations, with many sources classified in the star-forming/ULIRG region of WISE colour space. Note, only sources with a S/N $> 3$ in $\rm{W}1$, $\rm{W}2$, and $\rm{W}3$ are shown.}
    \label{sf_wmag_comparison}
\end{figure}

The distributions of WISE colour classifications for each source-finder are shown in Figure~\ref{sf_wmag_comparison}, with corresponding percentages in Table~\ref{sf_wmag_percentages}. Our sample contains a slightly higher fraction of star-forming hosts than reported by \citet{Gordon_DRAGN_2023} and \citet{Gupta_rgcat_2024}. Our proportion of star-forming hosts are however, consistent with the findings of \citet{mingo_agndiagnostic_2016} and \citet{mingo_revisiting_2019}, determined for classical FR-type and bent radio sources. The infrared hosts identified for CG-Complexity sources show the largest AGN fraction, with $\sim\!60\%$ lying in the AGN region, and the smallest proportion of star-forming hosts, with only $11.8\%$. This may reflect CG-Complexity’s capacity to identify sources at higher redshift. \textit{DH} and RG-CAT show similar colour distributions, each with $41$–$49\%$ in the AGN and $19$–$23\%$ star-forming regions, along with comparable ULIRG fractions. Passive galaxies comprise only a small minority ($4.7$–$8.7\%$) across all source-finders.

\begin{table}[h!]
\centering
\caption{The top section highlights the percentage of sources detected in AllWISE with S/N in $\rm{W}1$, $\rm{W}2$ and $\rm{W}3>3$, and those that are unclassified (either insufficient S/N or no detection in one or more bands), relative to the total number of AllWISE matches (shown under each source-finder, respectively). The bottom section shows the percentages of WISE colour classifications (AGN, Passive, Star-Forming, U-LIRG) for the S/N $>3$ subset, relative to the number of sources in the S/N $>3$ subset (shown under each source-finder, respectively). For all percentages, we include $1\sigma$ binomial uncertainties.}
\resizebox{\linewidth}{!}{%
\begin{tabular}{lcccc}
\toprule
 & \multicolumn{4}{c}{\textbf{Sources Detected in AllWISE}} \\
\cmidrule(lr){2-5}
Source-Finder & \multicolumn{2}{c}{S/N $>3$ (\%)} & \multicolumn{2}{c}{Unclassified (\%)} \\
\midrule
DRAGN{\footnotesize HUNTER} & \multicolumn{2}{c}{$18.48 \pm 0.93$} & \multicolumn{2}{c}{$81.52 \pm 0.93$} \\
(AllWISE $=1748$) & & & & \\
CG-Complexity & \multicolumn{2}{c}{$16.95 \pm 0.86$} & \multicolumn{2}{c}{$83.05 \pm 0.86$} \\
(AllWISE $=1900$) & & & & \\
RG-CAT & \multicolumn{2}{c}{$12.26 \pm 1.08$} & \multicolumn{2}{c}{$87.74 \pm 1.08$} \\
(AllWISE $=922$) & & & & \\
\midrule
 & \multicolumn{4}{c}{\textbf{WISE Colour Classifications}} \\
\cmidrule(lr){2-5}
Source-Finder & AGN (\%) & Passive (\%) & Star-Forming (\%) & ULIRG (\%) \\
\midrule
DRAGN{\footnotesize HUNTER} & $41.49 \pm 2.74$ & $8.98 \pm 1.59$ & $19.50 \pm 2.20$ & $30.03 \pm 2.55$ \\
($N=323$) & & & & \\
CG-Complexity & $61.80 \pm 2.71$ & $4.66 \pm 1.17$ & $11.80 \pm 1.80$ & $21.43 \pm 2.29$ \\
($N=322$) & & & & \\
RG-CAT & $49.56 \pm 4.70$ & $5.31 \pm 2.11$ & $23.01 \pm 3.96$ & $22.12 \pm 3.90$ \\
 ($N=113$) & & & & \\
\bottomrule
\end{tabular}}
\label{sf_wmag_percentages}
\end{table}

\begin{figure}[h!]
    \centering
    \includegraphics[width=\linewidth]{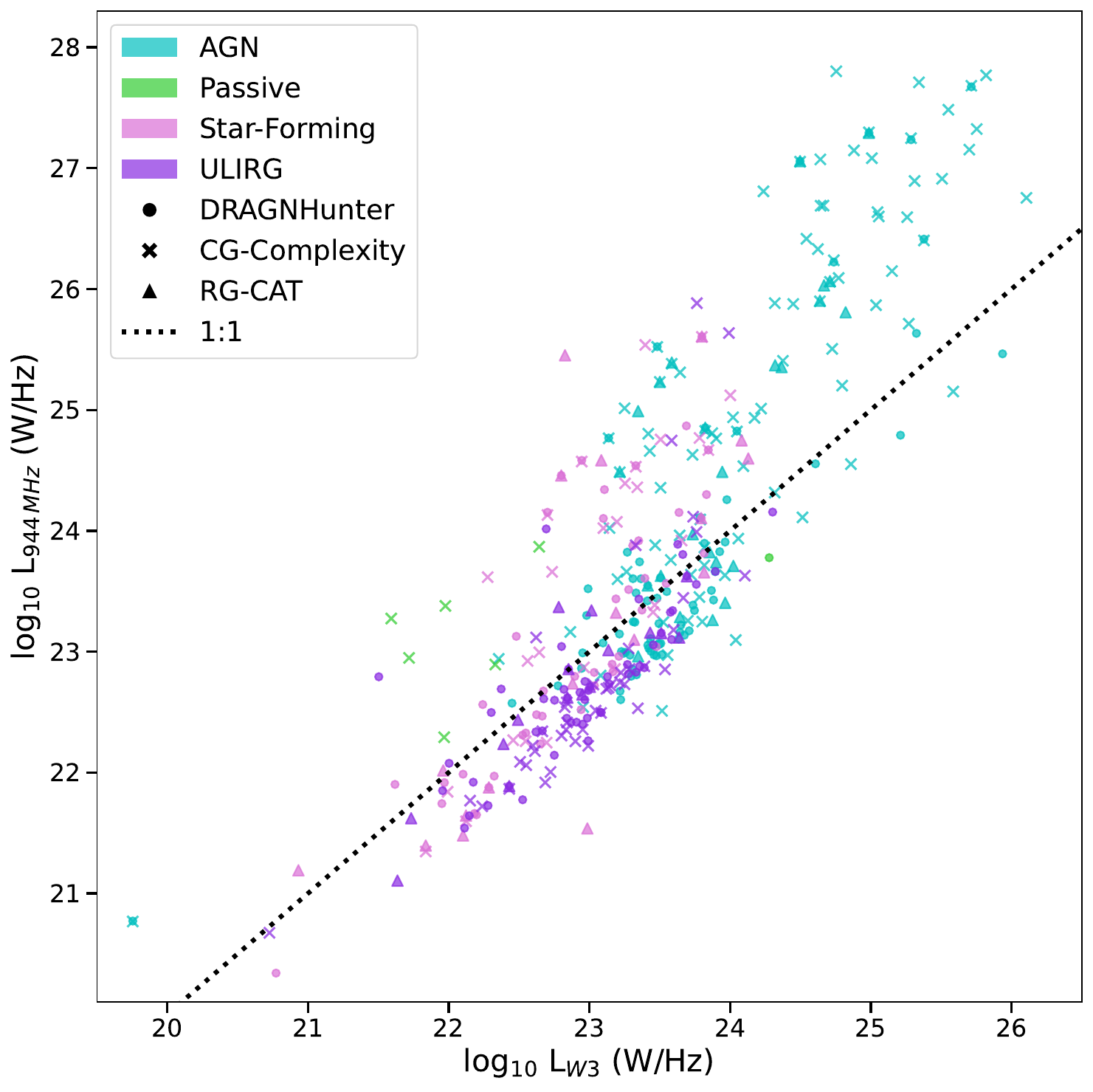}
    \caption{Radio luminosity versus $\rm{W}3$ luminosity for sources identified by each source-finder, colour–coded by WISE-colour classification (passive, star-forming, ULIRG, AGN). The dashed line shows the 1:1 reference, included as a visual guide only and not representing the true radio–IR correlation (better traced by the distribution of star-forming sources). Most sources lie above the line, indicating a radio excess consistent with AGN-dominated emission, even for hosts classified as star-forming. Only sources with S/N $> 3$ in $\rm{W}1$, $\rm{W}2$, and $\rm{W}3$ are shown.}
    \label{radlum_irlum}
\end{figure}

To further investigate the proportions from the WISE-colour classifications in our sample, we compare the radio luminosity and $\rm{W}3$ luminosity (Figure~\ref{radlum_irlum}). Most sources lie above the 1:1 reference line, indicating a radio excess relative to their mid-infrared emission, except for the majority of ULIRG-classified sources. Many of the WISE-AGN sources are identified by CG-Complexity, consistent with the idea that it detects compact or marginally resolved (potentially high redshift) sources. A small number of sources lie close to the distribution of star-forming galaxies, suggesting that some are genuinely powered by star formation. However, the overall population is dominated by systems showing significant radio excess, consistent with AGN-driven radio emission.

\begin{figure}
    \centering
    \includegraphics[width=\linewidth]{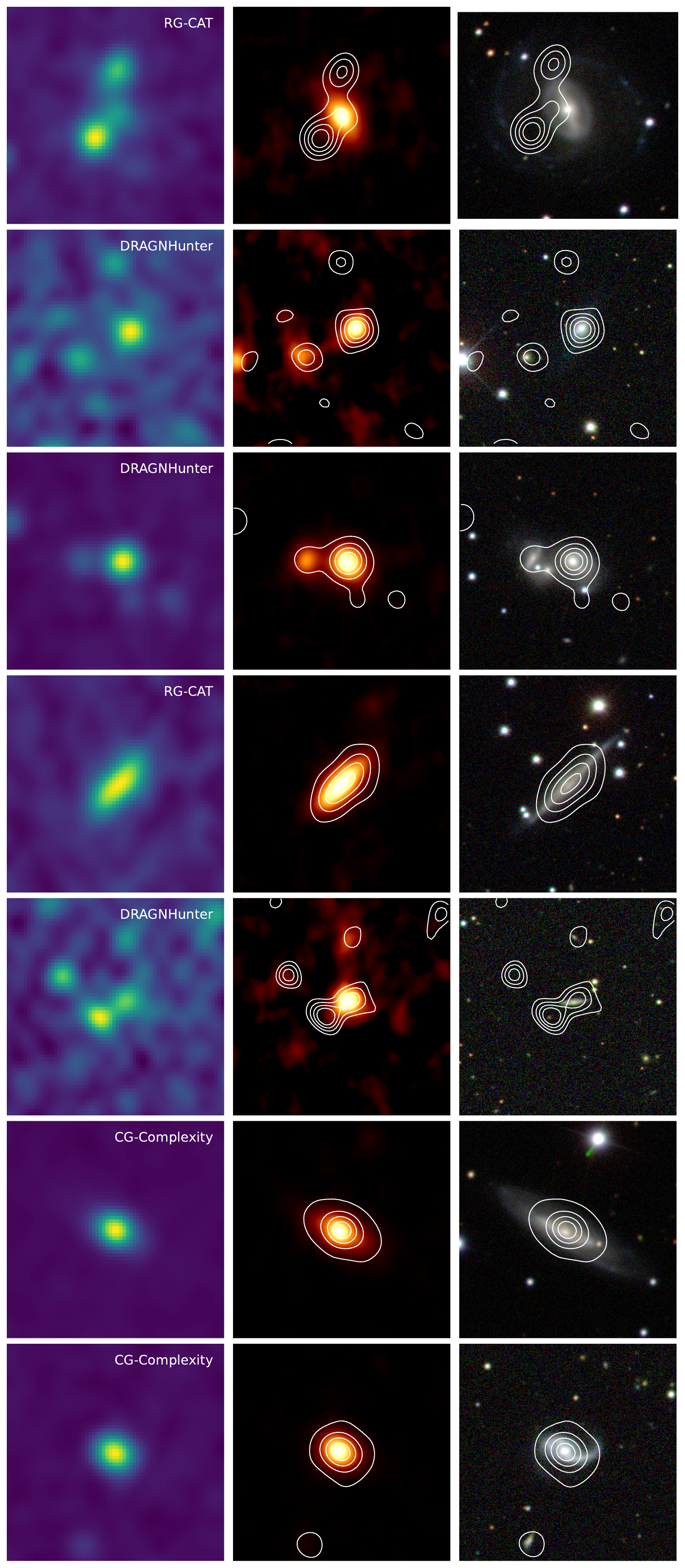}
    \caption{Example sources classified as star-forming or ULIRG based on their WISE colours. For each system, the left panel shows the radio emission with the detecting source-finder indicated in the top right, the middle panel shows the WISE $\mathrm{W3}$ ($12~\mu$m) image, and the right panel shows the corresponding SDSS composite-colour image using the \textit{g}, \textit{r}, \textit{i} bands. Contours in both the WISE and SDSS images correspond to the radio flux. All cutouts have an angular size of $128''\times128''$. Many of the WISE star-forming classifications appear to be from a resolved spiral galaxy with the radio emission coming from star-forming regions in the spiral arms.}
    \label{3bandcolor}
\end{figure}

We visually inspected sources classified as star-forming or ULIRG using WISE and SDSS images (Figure~\ref{3bandcolor}). For many of these sources, the SDSS composite image reveals a resolved/semi-resolved spiral galaxy, with the radio emission likely coming from HII star forming regions in the spiral arms. Interestingly, CG-Complexity has the smallest number of star-forming classifications, despite its capacity to detect irregular emission. It appears for \textit{DH} and RG-CAT, that two or more nearby spiral galaxies, with radio bright cores at their centre, can confuse the algorithms into a false detection of a radio double candidate.

\section{Conclusions}

The identification of extended radio sources remains a central challenge in large-area surveys, where it can be difficult to identify and characterise the complex morphology of radio sources by conventional source-finding algorithms. In this work, we applied the extended radio source-finders DRAGN{\footnotesize HUNTER}, coarse-grained complexity, and RG-CAT to EMU observations of the G09 field. 

We find that each source-finder identifies a largely independent set of sources, with only $375$ detections common to all three catalogues. This demonstrates that the choice of source-finding algorithm can substantially shape the resulting sample and highlights the absence of a single, universally optimal method for complex radio emission.

Despite this limited overlap, the identified sources share comparable physical properties. In particular, DRAGN{\scriptsize HUNTER} and RG-CAT recover sources of similar angular and linear sizes, flux densities, and luminosities, suggesting that these algorithms are sampling comparable astrophysical populations. Their infrared colour distributions are also nearly identical, indicating that all three methods trace host galaxies with similar underlying characteristics. Notably, a large proportion of the infrared-matched hosts fall outside canonical AGN regions, implying that many of the radio bright sources in EMU-G09 may not be powered by an AGN.

These results collectively show that no single source-finder provides a complete census of extended radio sources in the EMU-G09 field. Each approach contributes complementary detections that, when combined, yield a more comprehensive view of the sky. The similarities in the astrophysical properties across finders suggest that the physical interpretation of the detected sources is robust, even though any individual catalogue will be incomplete. Future work should therefore focus on integrating the strengths of these approaches, potentially through ensemble or hybrid methods, to ensure both completeness and reliability in upcoming large-scale surveys.

\begin{acknowledgement}
This scientific work uses data obtained from Inyarrimanha Ilgari Bundara, the CSIRO Murchison Radio-astronomy Observatory. We acknowledge the Wajarri Yamaji People as the Traditional Owners and native title holders of the Observatory site. CSIRO’s ASKAP radio telescope is part of the Australia Telescope National Facility (\url{https://ror.org/05qajvd42}). Operation of ASKAP is funded by the Australian Government with support from the National Collaborative Research Infrastructure Strategy. ASKAP uses the resources of the Pawsey Supercomputing Research Centre. Establishment of ASKAP, Inyarrimanha Ilgari Bundara, the CSIRO Murchison Radio-astronomy Observatory and the Pawsey Supercomputing Research Centre are initiatives of the Australian Government, with support from the Government of Western Australia and the Science and Industry Endowment Fund. This paper includes archived data obtained through the CSIRO ASKAP Science Data Archive, CASDA (\url{http://data.csiro.au}).

This publication makes use of data products from the Wide-field Infrared Survey Explorer, which is a joint project of the University of California, Los Angeles, and the Jet Propulsion Laboratory/California Institute of Technology, funded by the National Aeronautics and Space Administration. 

GAMA is a joint European-Australasian project based around a spectroscopic campaign using the Anglo-Australian Telescope. The GAMA input catalogue is based on data taken from the Sloan Digital Sky Survey and the UKIRT Infrared Deep Sky Survey. Complementary imaging of the GAMA regions is being obtained by a number of independent survey programmes including GALEX MIS, VST KiDS, VISTA VIKING, WISE, Herschel-ATLAS, GMRT and ASKAP providing UV to radio coverage. GAMA is funded by the STFC (UK), the ARC (Australia), the AAO, and the participating institutions. The GAMA website is \url{https://www.gama-survey.org/}.

L.J.B acknowledges the financial support provided by the Macquarie Research Excellence Scholarship (MQRES) program, received throughout the duration of this research.

We thank Denis Leahy and Michael Cowley for providing constructive comments on this paper.

\end{acknowledgement}

\printendnotes

\bibliography{main}

\appendix

\section{Small Pair Separation}\label{low_pair_sep}

\begin{figure}[h!]
    \centering
    \includegraphics[width=0.95\linewidth]{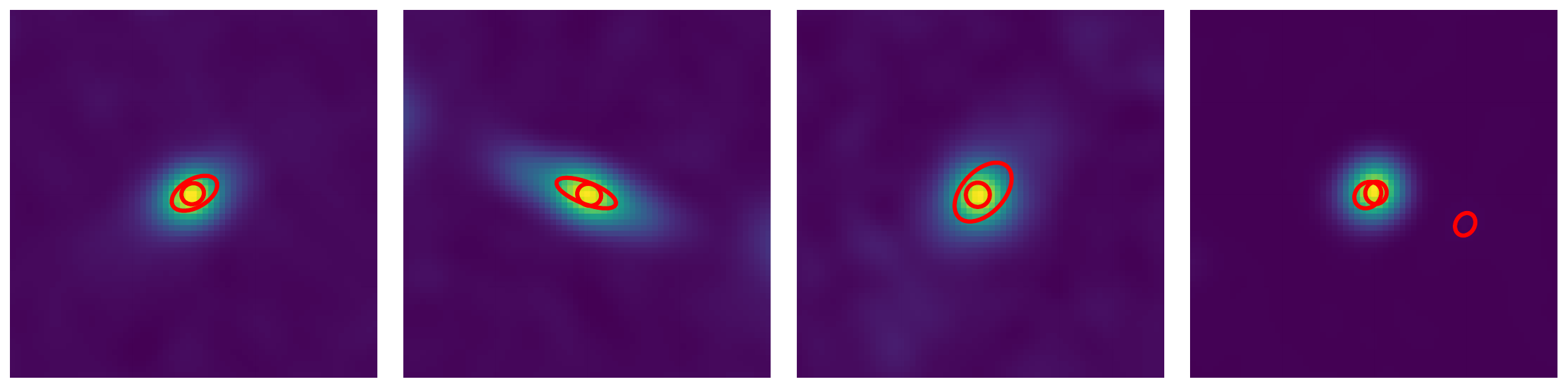}
    \caption{Sources with the lowest component pair separation. Overplotted are ellipses corresponding to the Gaussian components fit by \textit{Selavy}. For each of these sources, the components are nested, therefore producing extremely small component pair separation. All cutouts are $64\times64$ pixels, corresponding to an angular size of $128''\times128''$.}
    \label{min_sep_pair_sources}
\end{figure}

We find that $210$ or the candidate DRAGN{\footnotesize HUNTER} component pairs have an angular separation $<15''$ ($0.6\%$), with a minimum component separation of $0.659''$. In Figure~\ref{min_sep_pair_sources}, we highlight the four sources with the smallest pair separations. These cases are the result of multiple Gaussian components fit to a single island (e.g., a narrow Gaussian on the bright core and a broad Gaussian for the diffuse wing) that are nested within the same island.

\section{Zero Component Islands}\label{0_comp_islands}

\begin{figure}[h!]
    \centering
    \includegraphics[width=\linewidth]{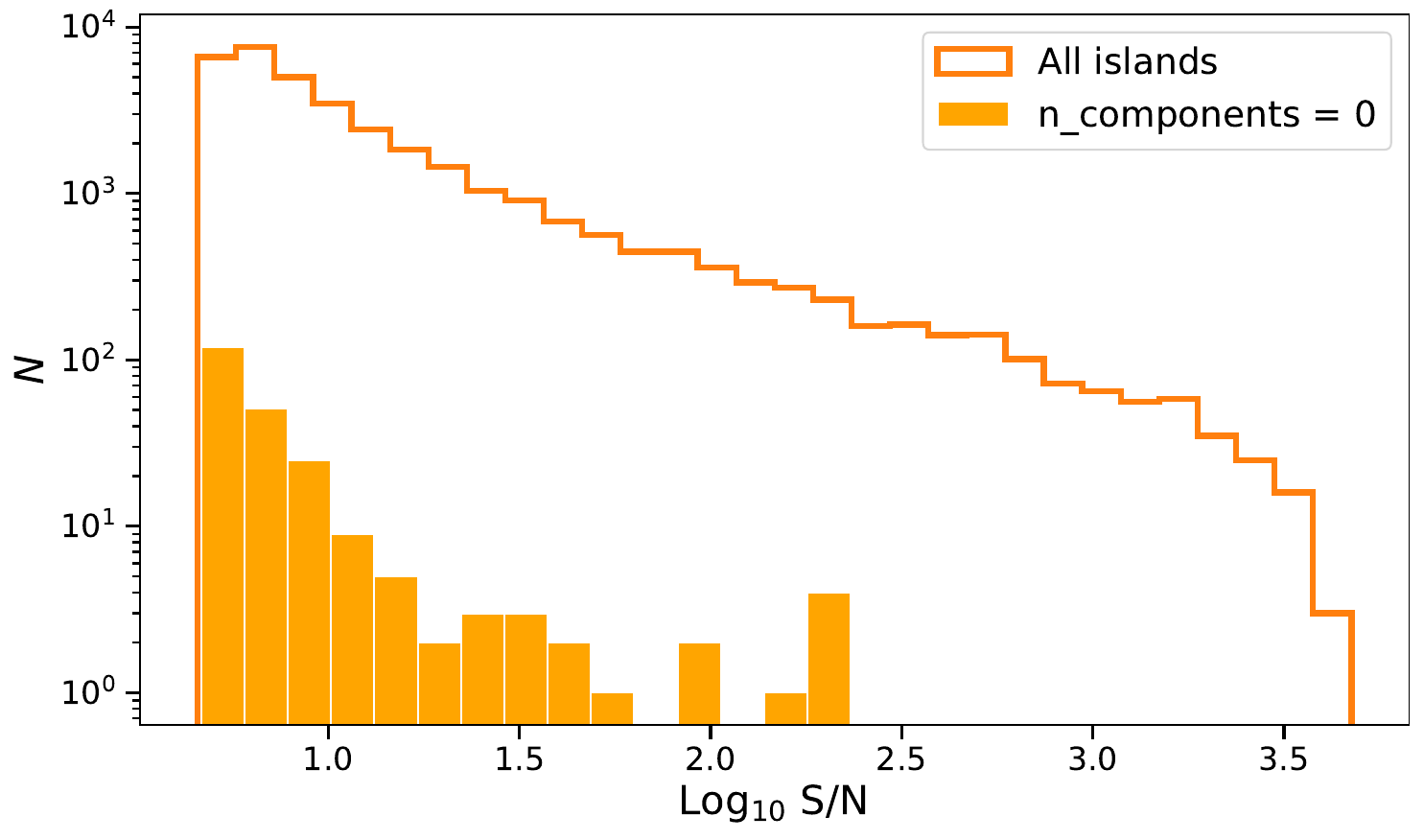}
    \caption{Signal to noise distributions of all islands (dark orange line) and islands with zero fitted components (orange bars).}
    \label{nocomp_snr}
\end{figure}

We find $228$ islands that were not fit with a Gaussian component by \textit{Selavy}. To further understand this, we first compare the distribution of signal-to-noise (S/N) of the islands with zero components to the entire sample (Figure~\ref{nocomp_snr}). The zero component islands tend to have low to moderate S/N, when compared to the rest of the islands.

\begin{figure}[h!]
    \centering
    \includegraphics[width=\linewidth]{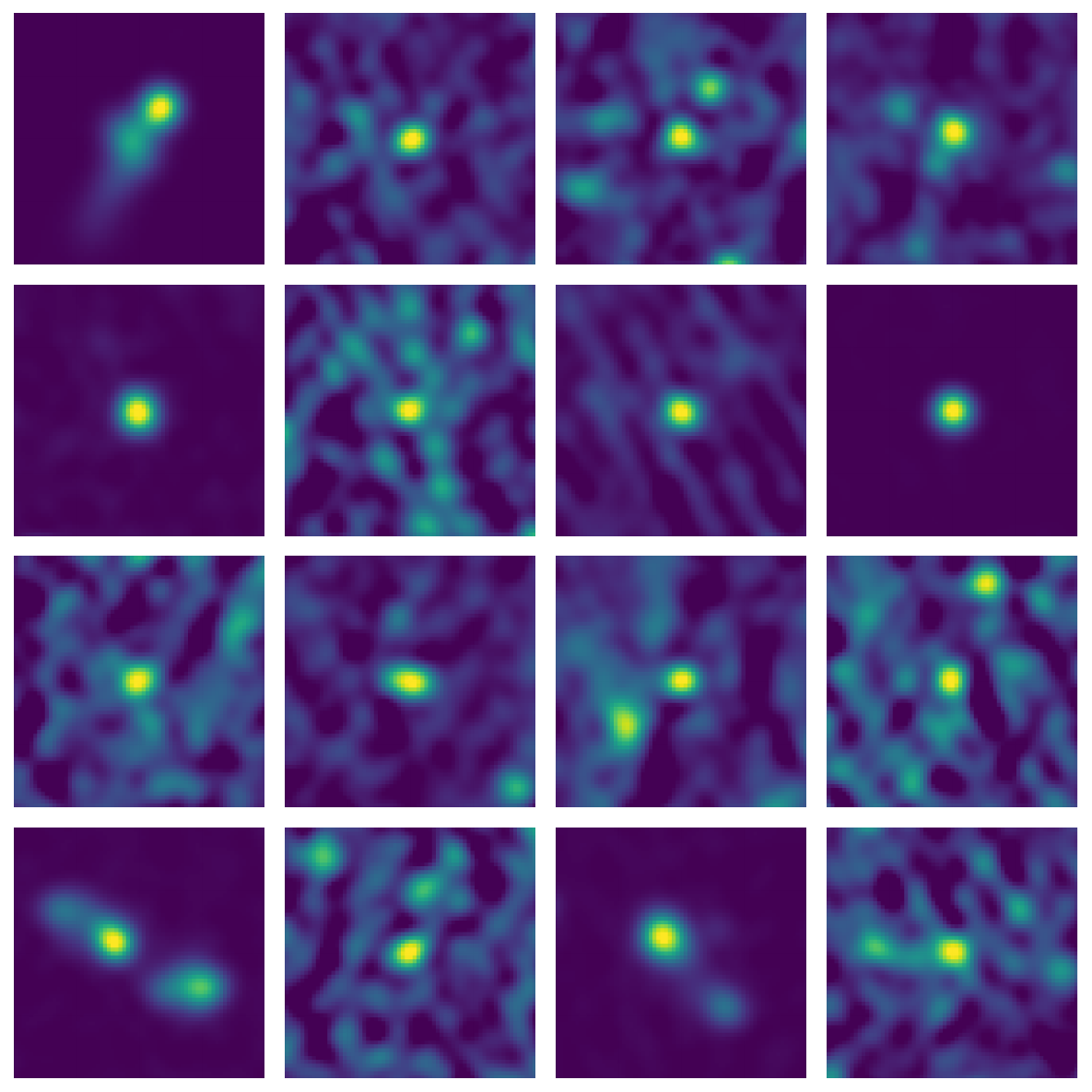}
    \caption{Cutouts of a random selection of 16 islands with zero fitted Gaussian components. All cutouts are $64\times64$ pixels, corresponding to an angular size of $128''\times128''$.}
    \label{nocomp_imgs}
\end{figure}

We then visually inspected the zero component sample. Figure~\ref{nocomp_imgs} shows a random selection of 16 images of islands with zero Gaussian components. This sample includes several low S/N sources which may prevent \textit{Selavy} from fitting Gaussian components. However, there are some brighter point sources, and some morphologically complex sources present. This suggests that the absence of fitted components does not necessarily indicate poor data quality. We therefore retain the zero-component islands in our analysis. The low number of zero-component sources, relative to the high complexity sample, should not impact our analysis. Their inclusion may also partly explain why the CG-Complexity approach identifies a larger number of sources than both \textit{DH} and RG-CAT.  

\section{`Complex' Point Sources}\label{CG_complex_points}

\begin{figure}[h!]
    \centering
    \includegraphics[width=\linewidth]{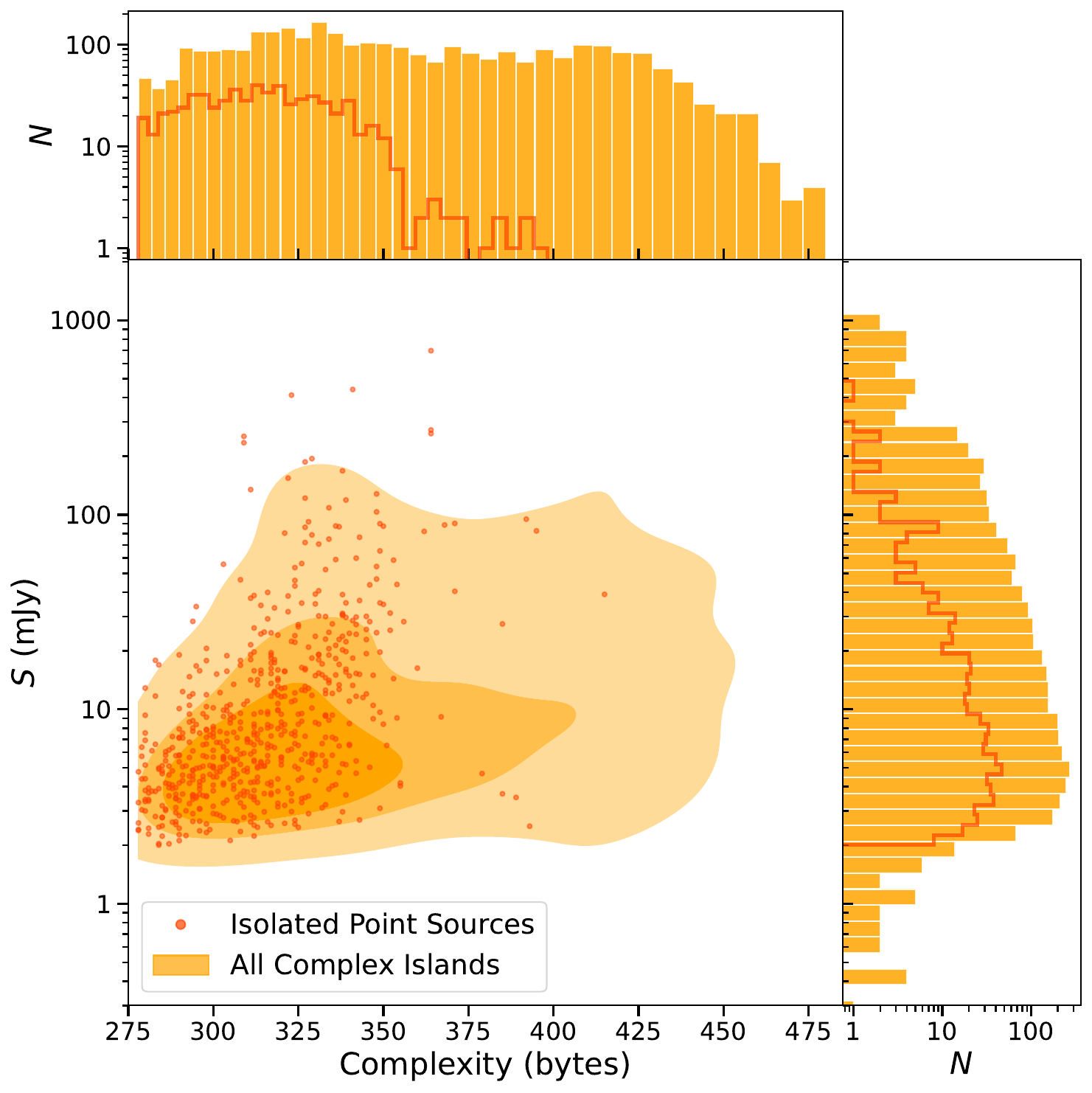}
    \caption{Distributions of complexity and flux density ($S$) for the entire high complexity sample (orange) and isolated point sources within the high complexity sample (dark orange). Density contours of the distributions of complexity and $S$ for the full high complexity sample contain $90\%$, $50\%$, and $25\%$ of the data points.}
    \label{cg_point_sources}
\end{figure}

To investigate why point sources appear in the high-complexity subset, we cross-matched islands fitted with a single Gaussian component to the \textit{Selavy} component catalogue in order to obtain their angular sizes. We then defined a threshold below which single-component islands are considered unresolved. The minimum component size among single-component islands is $14.39''$. A visual inspection with different thresholds indicated that a threshold of $17''$ effectively separates the majority of genuinely extended sources in the high-complexity sample from sources that appear to be unresolved. Sources where the fitted component is below this $17''$ are considered point sources. We then performed a nearest-neighbour search on the point source sample using a radius that captures the full cutout size employed in the complexity calculation (see \S~\ref{cg_comp_section}). Sources with no neighbours within this radius are considered isolated point sources. We further explore the isolated point sources by examining their complexity values and flux density, relative to the full highly complex sample in Figure~\ref{cg_point_sources}.

We find that the isolated point sources generally occupy the lower end of the complexity distribution relative to the full high-complexity sample, while exhibiting a broadly similar flux density distribution. This suggests that their inclusion in the high-complexity subset is not driven purely by extreme brightness. Instead, the results are consistent with a weakly constrained relationship between apparent complexity and flux density, whereby sufficiently bright compact sources, even when isolated, could be deemed as `complex' by the CG-Complexity measurement. This behaviour may partly reflect the signal-to-noise criterion incorporated in the second GMM used to define the highly complex sample: intrinsically bright sources naturally exhibit high S/N and may therefore be preferentially included, even when their underlying morphology is simple.

If the goal is to minimise contamination from isolated point sources prior to the complexity analysis, the procedure outlined above (component matching, application of a size threshold, and a nearest-neighbour isolation check) can be applied before running the two-stage GMM selection. This would preferentially remove compact, isolated sources that may be retained after the two step GMM filtering. In this work, however, we deliberately retain these sources to preserve a minimally tuned implementation of the CG-Complexity pipeline and to provide a representative assessment of the intrinsic behaviour of each source-finding approach under broadly consistent conditions.

\section{\textit{DH} Likelihood Ratio Calculation}\label{DH_LR}

The full description of how the likelihood ratio (LR) is determined in the \textit{DH} script is detailed in \S4.2 of \citet{Gordon_DRAGN_2023}. For brevity, we only include the most relevant equations from \citet{Gordon_DRAGN_2023}. LR is defined as: 

\begin{equation}
    \text{LR} = \frac{q(\rm{W}1)f(r)}{n(\rm{W}1)}~,
\end{equation}

\noindent where $q(\rm{W}1)$ is probability that the radio source has an AllWISE counterpart with a given magnitude in the WISE $\rm{W}1$-band, $f(r)$ is the radial separation probability distribution function for the cross match, and $n(\rm{W}1)$ is the sky distribution of AllWISE sources of a given $\rm{W}1$-band magnitude. Once LR has been determined for each host, the reliability of any given match, $R_i$, can then be determined for the match between the radio source and $i$th AllWISE candidate out of $N$ possible matches:

\begin{equation}
    R_i = \frac{LR_i}{\Sigma^{N}_{j=1} LR_j + (1-Q_0)}~,
\end{equation}

\noindent where $Q_0$ is an estimate of the fraction of radio sources with an AllWISE match. The \textit{DH} script only adopts an AllWISE host with $R>0.5$. 

\begin{figure}[h!]
    \centering
    \includegraphics[width=\linewidth]{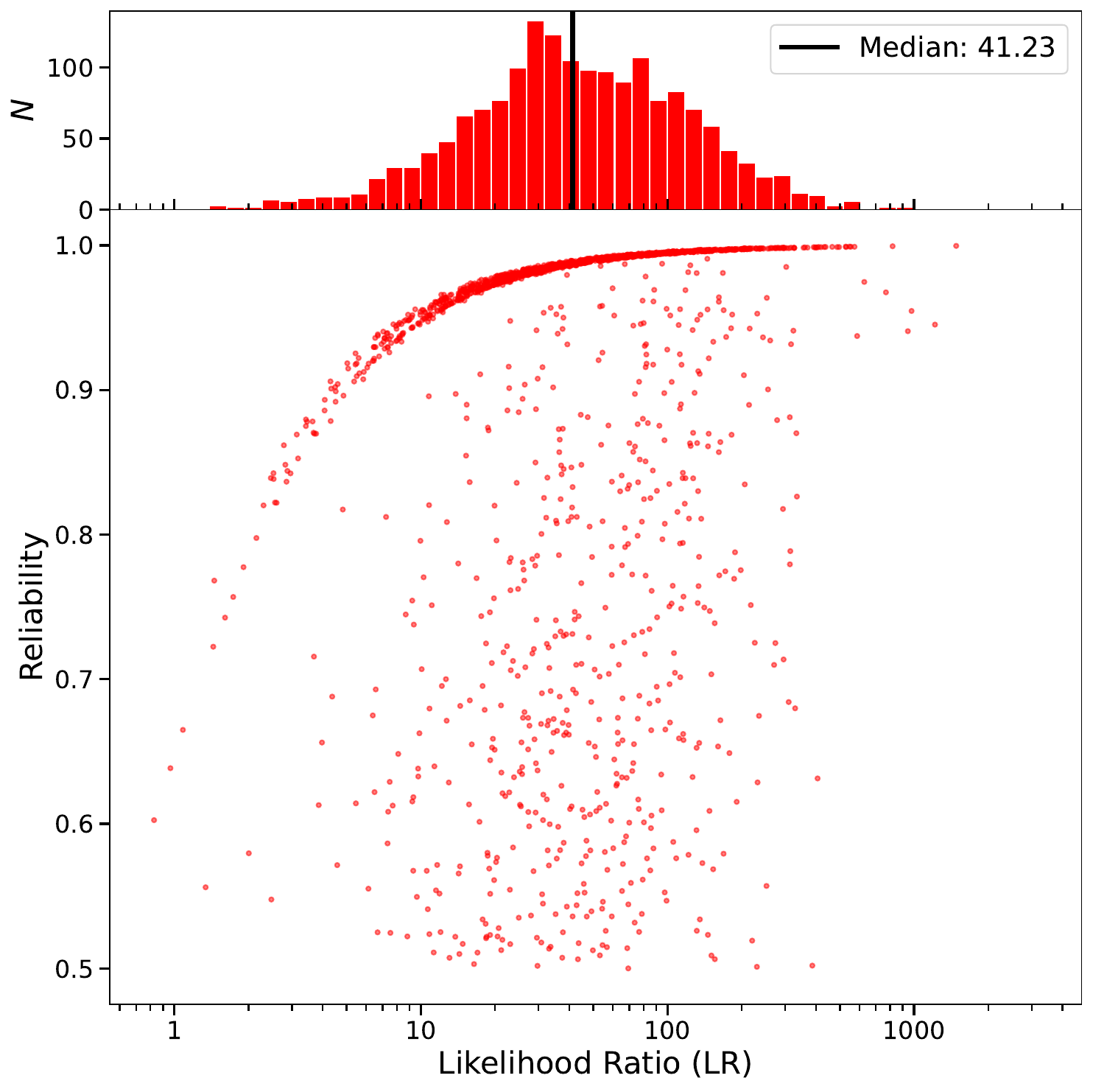}
    \caption{Top panel: Distribution of likelihood ratio values for \textit{DH} sources crossmatched with AllWISE sources. Bottom panel: Likelihood ratio vs the calculated reliability for each cross-match.}
    \label{dh_lr}
\end{figure}

Figure~\ref{dh_lr} shows both the distribution of LR values and the relationship between LR and reliability for all \textit{DH} sources matched with AllWISE. The upper panel highlights that the majority of matches have a low to moderate LR (median of $41.23$). This is a significantly lower value than the LR peak of the VLASS data (peaking closer to $\sim\!1000$). This indicates that the cross-matching for the EMU sources is less reliable than the cross-matching for the VLASS sources, likely due to the resolution and sensitivity differences between surveys. We find a mean reliability ($R$) of $0.9$, corresponding to an estimated false positive rate (fpr, calculated by $1 - R$) for \textit{DH}-AllWISE cross-matches of approximately $\sim\!10.2\%$.

\end{document}